\documentclass{sig-alternate}

\usepackage{epsfig,endnotes}

\usepackage{subfig}
\usepackage[export]{adjustbox}
\usepackage{amsmath,lipsum}
\usepackage{listings}
\usepackage[table]{xcolor}
\usepackage{multirow}
\usepackage{enumitem}
\usepackage[hyphens]{url}
\usepackage{algorithm}
\usepackage[noend]{algpseudocode}
\usepackage{stfloats}
\usepackage{balance}
\usepackage{microtype}

\usepackage{tikz}
 
\usetikzlibrary{decorations.pathreplacing}
\usetikzlibrary{positioning}
\usetikzlibrary{patterns,shapes,arrows}
\usetikzlibrary{fit,calc}
\pgfdeclarelayer{background}
\pgfdeclarelayer{foreground}
\pgfsetlayers{background,main,foreground}

\algloop{continue}
\algloop{given}

\usepackage{xspace}
\usepackage{xparse}
\usepackage{setspace}
\newcommand{\hilite}[1]{{\setlength{\fboxsep}{1pt}\colorbox{magenta!30}{#1}}}
\newcommand{\white}[1]{{\setlength{\fboxsep}{1pt}\colorbox{magenta!0}{#1}}}
 
\lstdefinelanguage{Cmorekeywords}{
  language = C,
  morekeywords = {
    uint8\_t,
    uint16\_t,
    time_t,
  }
}
\lstnewenvironment{Ccode}
                  {\lstset{language = Cmorekeywords}}
                  {}

\newcommand{\hex}[1]{%
  \texttt{\textbackslash x}\texttt{#1}%
}

\usepackage[square,numbers,sectionbib]{natbib}
\renewcommand{\bibsection}{\section*{REFERENCES}}

\DeclareRobustCommand{\ttfamily}{\fontencoding{T1}\fontfamily{lmtt}\selectfont}

\begin{document}

\date{}

\title{Limitless HTTP in an HTTPS World: Inferring the Semantics of the HTTPS Protocol without Decryption}


\numberofauthors{4}
\author{
\alignauthor Blake Anderson \\
       \affaddr{Cisco Systems, Inc.} \\
       \email{blaander@cisco.com}\\
\alignauthor Andrew Chi \\
       \affaddr{University of North Carolina} \\
       \email{achi@cs.unc.edu}
\alignauthor Scott Dunlop \\
       \affaddr{Cisco Systems, Inc.} \\
       \email{scdunlop@cisco.com}
\and
\alignauthor David McGrew \\
       \affaddr{Cisco Systems, Inc.} \\
       \email{mcgrew@cisco.com}
}

\maketitle


\begin{abstract}
We present new analytic techniques for inferring HTTP semantics from passive observations of HTTPS that can infer the value of important fields including the \texttt{status-code}, \texttt{Content-Type}, and \texttt{Server}, and the presence or absence of several additional HTTP header fields, e.g., \texttt{Cookie} and \texttt{Referer}. Our goals are twofold: to better understand the limitations of the confidentiality of HTTPS, and to explore benign uses of traffic analysis such as application troubleshooting and malware detection that could replace HTTPS interception and static private keys in some scenarios. We found that our techniques improve the efficacy of malware detection, but they do not enable more powerful website fingerprinting attacks against Tor. Our broader set of results raises concerns about the confidentiality goals of TLS relative to a user's expectation of privacy, warranting future research.

We apply our methods to the semantics of both HTTP/1.1 and HTTP/2 on data collected from automated runs of Firefox 58.0, Chrome 63.0, and Tor Browser 7.0.11 in a lab setting, and from applications running in a malware sandbox. We obtain ground truth plaintext for a diverse set of applications from the malware sandbox by extracting the key material needed for decryption from RAM post-execution. We developed an iterative approach to simultaneously solve several multi-class (field values) and binary (field presence) classification problems, and we show that our inference algorithm achieves an unweighted $F_1$ score greater than 0.900 for most HTTP fields examined.

\end{abstract}


\section{Introduction}

\noindent HTTPS, or HTTP-over-TLS, encrypts HTTP requests and responses, and is foundational to internet security. In this paper, we show that it is possible to infer the HTTP method, status-code, and header fields without decrypting the connection. Inferring HTTP protocol semantics from observations of HTTPS connections aims to provide ``Limitless HTTP in an HTTPS World", to adapt an ironic phrase\footnote{Michael Scott's unwittingly incongruous slogan from \textit{The Office}, ``Limitless Paper in a Paperless World".}. A primary goal for TLS is confidentiality \cite{tls12, tls13}; our work explores places where it fails due to the gap between theory and practice. Semantic security \cite{Goldwasser:1982:PEA:800070.802212, DBLP:journals/jcss/GoldwasserM84} requires that an attacker has negligible advantage in guessing any plaintext value, but the cryptographic algorithms in TLS only meet the lesser goal of indistinguishability of the plaintext from messages of the same size. A cipher meeting the lesser goal fails at semantic security when applied to the message space consisting of the strings \texttt{YES} and \texttt{NO}. While our proposed methods cannot undo encryption entirely, they can often recover detailed information about the underlying HTTP protocol plaintext, which raises questions about what goals for HTTPS are most appropriate and achievable.  Many users would reject a hypothetical variant of HTTPS that left the method, content-type, and status codes unencrypted 98\% of the time, yet our analysis produces essentially the same outcome on standard HTTPS implementations.



We frame the HTTP protocol semantics inference problem as a number of disjoint multi-class and binary classification problems. The multi-class classifiers model field values, e.g., \texttt{nginx-1.13} for the \texttt{Server} field and \texttt{text/html} for the \texttt{Content-Type} field. The binary classifiers model the presence or absence of a field, e.g., the presence of the \texttt{Cookie} and \texttt{Referer} fields in an HTTP request. We designed classifiers for the \texttt{method} and \texttt{status-code} fields as well as HTTP headers that were well-represented in both our HTTP/1.1 and HTTP/2 datasets and exhibited a reasonable level of diversity, i.e., the majority class label appeared in less than 80\% of the decrypted sessions. Many of these values are correlated with other values in the same HTTP transaction or other transactions in the same TLS connection, e.g., \texttt{text/css} and \texttt{text/javascript} objects are often transferred using the same TLS connection. Using this intuition, we developed an iterative classification strategy that utilizes the header field inferences of related transactions that were predicted during the previous iteration.

Our HTTP inference experiments uses data collected from Firefox 58.0, Chrome 63.0, Tor Browser 7.0.11, and data collected from a malware sandbox over a two week period. Data collected in the first week was used for training and data collected in the second week was used for testing, which reflects how these methods would be used in practice and highlights the importance of maintaining a current training dataset. The malware sandbox data was generated by automated runs of submitted samples, and the browser data was generated daily by collecting all connections after initiating a connection to each website in the Alexa top-1,000 list.

These experiments were designed to test optimistic and more realistic deployments of our techniques. In the optimistic deployment, the splitting of the Firefox, Chrome, and Tor datasets into training and testing sets resulted in some overfitting. We believe these experiments are still informative because inferences capturing the purpose of common connections with high efficacy are valuable, and our data collection strategy captured dynamic content directly loaded by the target website, e.g., news stories, and indirectly loaded by the target website, e.g., referred advertising sites, resulting in a more temporally diverse dataset than one would expect. The HTTP inferences on the malware dataset reflected settings where the model must generalize further. The majority of the malware HTTP inferences' test dataset was composed of sites not present during training, and the malware exhibited far more application diversity

As with all traffic analysis research, our results have implications for both attackers and defenders; we touch on both sides of this dichotomy throughout this paper, and examine two popular research areas that embody the moral tension of our results. The first, malware detection, has become increasingly important with the rise of encrypted traffic, and malware's predictable use of encryption to obfuscate its network behavior \cite{anderson16identifying,anderson16deciphering}. The second, website fingerprinting, has serious privacy implications and is most often examined in the context of the Tor protocol in the literature. Our techniques did not improve the performance of website fingerprinting, which we attribute to Tor's use of fixed-length cells and multiplexing.  On the other hand, we found that HTTP protocol semantics inferences can improve the detection of TLS encrypted malware communication. We attribute this increase in performance to the differences in the distributions of HTTP header fields, e.g., requested \texttt{Content-Type} and \texttt{Server} fields, and the presentation of these learned concepts to the malware classifier.


HTTPS traffic is often subject to interception to detect and block malicious content \cite{httpsInterception2017} and to passive monitoring using static private keys \cite{draft-green-tls-static-dh-in-tls13-01}.  Benign traffic analysis offers an alternative that better respects the principle of least privilege. Green et al. \cite{draft-green-tls-static-dh-in-tls13-01} cite application troubleshooting and performance analysis as major motivations for passive HTTPS monitoring, and our HTTP inferences can be directly applied to those use cases, without third-party decryption or key escrow. Similarly, benign traffic analysis may allow some network administrators to avoid the use of TLS termination proxies and the associated security issues  \cite{httpsInterception2017,uscert2017https}. This approach has many advantages over actively probing servers. Specifically, active probes have difficulty accounting for all possible client configurations, the server and associated software need to be active during the scan, and the probes will not necessarily exercise problematic server options. For example, 3.2\% of the HTTP/2 connections we observed used multiple web servers within the same connection, and this behavior was often dependent on the requested \texttt{URI}. These connections typically included proxy servers, servers providing static and dynamic content, and servers processing client data. Examples of this behavior include YouTube's \texttt{YouTube Frontend Proxy}/\texttt{sffe}/\texttt{ESF} stack and CNN's \texttt{Akamai Resource Optimizer}/\texttt{Apache}/\texttt{Apache-Coyote-1.1}/\texttt{nginx} stack.


To implement the algorithms outlined in this paper, the adversary or defender is assumed to have several capabilities. First, they need the ability to passively monitor some target network's traffic. Second, they need to have a large, diverse, and current corpus of training data correlating the encrypted traffic patterns observable on a network with the underlying HTTP transactions. Third, the adversary or defender needs to have the computational power to execute many classifiers per observed TLS connection.

We make the following novel contributions:
\begin{enumerate}[itemsep=-1mm]
\item We describe the first framework to infer an extensive set of HTTP/1.1 and HTTP/2 protocol semantics inside of TLS and Tor encrypted tunnels without performing decryption.
\item We test our algorithms on datasets based on Firefox 58.0, Chrome 63.0, Tor Browser 7.0.11, and data collected from a malware sandbox. We show that we can reliably infer the semantics of HTTP messages on all datasets except for Tor.
\item We provide the community with an open sourced dataset containing the packet captures and the cryptographic key material for the Firefox 58.0, Chrome 63.0, and Tor Browser 7.0.11 datasets \footnote{Release is forthcoming.}.
\item We apply our methods to TLS encrypted malware detection and Tor website fingerprinting. We show that first modeling the semantics of encrypted HTTP messages has the potential to improve malware detection, but fails to improve website fingerprinting due to Tor's implemented defenses.
\end{enumerate}

\tikzset{ablock/.style={draw, text width=3.2cm, minimum height=13pt, align=center, node distance=.3cm, fill=white},
         bblock/.style={draw, text width=1.8cm, minimum height=13pt, align=center, node distance=.3cm, fill=white},
         cblock/.style={draw, text width=5cm, minimum height=13pt, align=center, node distance=.3cm, fill=white},
         dblock/.style={draw, text width=9cm, minimum height=13pt, align=center, node distance=.3cm, fill=white},
         null/.style={rectangle, minimum height=13pt, minimum width=0pt, node distance=.25cm},
  line/.style = {-latex}
}
\begin{figure*}[thb!]
\centering 
\resizebox{130mm}{!}{
\begin{tikzpicture}

\node at (-1,.8) {\textbf Client};
\node at (14.660,.8) {\textbf Server};

\draw[dashed, line width=.4mm] (-1.1,.5) -- (-1.1,-4.01);
\draw[dashed, line width=.4mm] (14.760,.5) -- (14.760,-4.01);

  \node[null]                              (a0) {};
  \node[bblock, right= of a0, fill=blue!10!white]              (a) { \footnotesize\ttfamily Preface };
  \node[bblock, right= of a, fill=blue!10!white]               (b) { \footnotesize\ttfamily SETTINGS };
  \node[ablock, right= of b, fill=blue!10!white]               (c) { \footnotesize\ttfamily WINDOW\_UPDATE };
  \node[bblock, right= of c, fill=blue!10!white]               (d) { \footnotesize\ttfamily PRIORITY };
  \node[bblock, right= of d, fill=blue!10!white]               (e) { \footnotesize\ttfamily PRIORITY };
  \node[null, right= of e]                 (f0) {};

  \node[null, below= of a0]                (aa0) {};
  \node[null, below= of f0]                (ff0) {};
  \node[cblock, right= of aa0, fill=blue!10!white]             (aa) { \footnotesize\ttfamily HEADERS };
  \node[bblock, right= of aa, fill=blue!10!white]              (bb) { \footnotesize\ttfamily PRIORITY };
  \node[null, below= of aa0]               (aaa0) {};
  \node[null, below= of ff0]               (fff0) {};
  \node[ablock, left= of fff0, fill=blue!10!white]             (eee) { \footnotesize\ttfamily WINDOW\_UPDATE };
  \node[bblock, left= of eee, fill=blue!10!white]              (ddd) { \footnotesize\ttfamily SETTINGS };
  \node[null, below= of aaa0]              (aaaa0) {};
  \node[null, below= of fff0]              (ffff0) {};
  \node[bblock, right= of aaaa0, fill=blue!10!white]           (bbbb) { \footnotesize\ttfamily SETTINGS };
  \node[null, below= of aaaa0]              (aaaaa0) {};
  \node[null, below= of ffff0]              (fffff0) {};
  \node[cblock, left= of fffff0, fill=blue!10!white]            (fffff) { \footnotesize\ttfamily HEADERS };
  \node[null, below= of aaaaa0]              (aaaaaa0) {};
  \node[null, below= of fffff0]              (ffffff0) {};
  \node[dblock, left= of ffffff0, fill=blue!10!white]            (ffffff) { \footnotesize\ttfamily DATA }; 

  \begin{pgfonlayer}{background}
    \draw[line width=.4mm, ->] (-1.1, 0.01) -- (14.760,0.01);
    \draw[line width=.4mm, ->] (-1.1, -.72) -- (14.760,-.72);
    \draw[line width=.4mm, <-] (-1.1,-1.45) -- (14.760,-1.45);
    \draw[line width=.4mm, ->] (-1.1,-2.17) -- (14.760,-2.17);
    \draw[line width=.4mm, <-] (-1.1,-2.89) -- (14.760,-2.89);
    \draw[line width=.4mm, <-] (-1.1,-3.61) -- (14.760,-3.61);

    \node[draw, inner sep=2pt, fit=(a) (b) (c) (d) (e), fill=black!40!white] {};
    \node[draw, inner sep=2pt, fit=(aa) (bb), fill=black!40!white]           {};
    \node[draw, inner sep=2pt, fit=(eee)     (ddd), fill=black!40!white]           {};
    \node[draw, inner sep=2pt, fit=(bbbb), fill=black!40!white]               {};
    \node[draw, inner sep=2pt, fit=(fffff), fill=black!40!white]               {};
    \node[draw, inner sep=2pt, fit=(ffffff), fill=black!40!white]               {}; 
  \end{pgfonlayer}

\end{tikzpicture}
}
\caption{Firefox 58.0 HTTP/2 connection to \texttt{google.com} inside of a TLS encrypted tunnel. The gray boxes indicate a single TLS \texttt{application\_data} record, and the light blue boxes indicate HTTP/2 frames.}
\label{fig:h2-session}
\end{figure*}
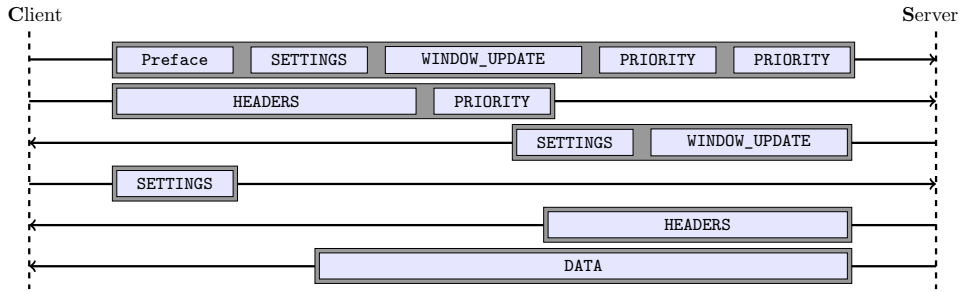

\section{Background}

\subsection{Relevant Protocols}

\noindent The data and analysis of this paper revolves around making inferences relative to 4 main protocols: HTTP/1.1 \cite{http11_message_syntax,http11_semantics_content}, HTTP/2 \cite{http2}, TLS 1.2 \cite{tls12}, and Tor \cite{tor}. Other Transport Layer Security (TLS) versions, such as TLS 1.3 \cite{tls13}, did appear in our data, but represented less than 5\% of connections.

HTTP is the prominent protocol to facilitate data transfer on the World Wide Web. HTTP/1.1 \cite{http11_message_syntax,http11_semantics_content} and HTTP/2 \cite{http2} are the most popular versions of the HTTP protocol, and as shown in Table \ref{table:dataset_summary}, their representation was roughly equal in our datasets. HTTP/1.1 is a stateless protocol that enables the exchange of requests and responses between a client and server. An HTTP/1.1 request begins with a \texttt{request-line} specifying the \texttt{method}, \texttt{request-target}, and \texttt{HTTP-version} of the request, and a response begins with a \texttt{status-line} specifying the \texttt{HTTP-version}, \texttt{status-code}, and \texttt{reason-phrase}. Following the \texttt{request-line} or \texttt{status-line}, there is a potentially unordered, case-insensitive list of header fields with their associated values. In this paper, we make inferences on the \texttt{request-line}'s \texttt{method}, the \texttt{status-line}'s \texttt{status-code}, and many of the headers fields and values, such as \texttt{Referer}, \texttt{Server}, and \texttt{Content-Type}. HTTP/1.1 supports pipelining, where a client can send 2 or more requests before receiving the server's response, and the server will then send a series of responses in the same order as they were requested.

HTTP/2 \cite{http2} was introduced to solve some of HTTP/1.1's shortcomings, e.g., by introducing multiplexing and header compression. HTTP/2 can multiplex multiple streams over a single HTTP/2 connection by using various frames to communicate the state of different streams. Figure \ref{fig:h2-session} illustrates the creation of an HTTP/2 connection inside of a TLS tunnel. The dark gray boxes represent TLS \texttt{application\_data} records, and the light blue boxes represent HTTP/2 frames. The client begins an HTTP/2 connection by sending a fixed set of bytes indicating the connection is HTTP/2 immediately followed by a \texttt{SETTINGS} frame containing the parameters related to the client's HTTP/2 configuration. The client can send additional frames at this point, and in Figure \ref{fig:h2-session}, the client sends a \texttt{WINDOW\_UPDATE} frame for flow control management, a set of \texttt{PRIORITY} frames defining the priority of the multiple streams in the connection, a \texttt{HEADERS} frame containing the request header fields and values, and finally another \texttt{PRIORITY} frame. The server must begin an HTTP/2 connection by sending a \texttt{SETTINGS} frame. After an additional exchange of \texttt{SETTINGS} frames, the server sends a \texttt{HEADERS} frame containing the response header fields and values, and finally a \texttt{DATA} frame containing the requested data. The header fields are compressed using HPACK \cite{hpack}. In our experiments, we are only concerned with identifying \texttt{HEADERS} frames and the values they contain.

HTTP/1.1 requests and responses are increasingly being secured by TLS, and browser vendors have stated that they will not implement HTTP/2 without encryption. The TLS handshake begins with the exchange of \texttt{client\_hello} and \texttt{server\_hello} records that establish the cryptographic parameters needed to encrypt and authenticate data. The client and server can also negotiate the application layer protocol with these messages by using the \texttt{application\_layer\_protocol\_negotiation} extension, where the values \texttt{http/1.1} and \texttt{h2} reasonably refer to HTTP/1.1 and HTTP/2. After establishing a set of shared keys, the client and server each send \texttt{change\_cipher\_spec} and \texttt{finished} records designating the end of the TLS handshake, and can now send encrypted \texttt{application\_data} records containing application layer data. All of the dark gray boxes in Figure \ref{fig:h2-session} represent \texttt{application\_data} records.

Tor securely transmits and anonymizes TCP-based application layer protocols, e.g., HTTP, using a combination of TLS, its own encryption protocol, and an overlay network \cite{tor}. The client creates a Tor connection by first negotiating a TLS handshake with a Tor entry node. After performing a Tor handshake, the client then constructs a circuit by sending a \texttt{CREATE2} cell to the first onion router in the chain, where a cell is the basic unit of communication similar to an HTTP/2 frame. The onion router responds with a \texttt{CREATED2} cell with the information needed to derive a pair of 128-bit AES keys to encrypt outgoing messages and decrypt incoming messages. The client sends an \texttt{RELAY\_EXTEND2} cell to extend the circuit to an additional onion router, and will follow the same key establishment protocol with the contents of the returned \texttt{RELAY\_EXTENDED2} cell. After repeating this process multiple times, \texttt{RELAY\_DATA} cells are sequentially encrypted with the 128-bit AES keys of each onion router in the circuit's path. The contents of \texttt{RELAY\_DATA} cells carry the relevant application layer data, e.g., TLS records needed to perform an additional TLS handshake, and are null padded so that they always contain 514 bytes. Figure \ref{fig:tor-json} shows a JSON representation of an HTTP/1.1 \texttt{GET} request tunneled over TLS, tunneled over Tor, and again tunneled over TLS.

\subsection{Inferences on Encrypted Traffic}

\noindent Inferring the content and intent of encrypted traffic has a rich history in the academic literature. While not directly targeting encrypted traffic, protocol-agnostic network threat detection can be applied to encrypted communication with reasonable results. These methods rely on data features such as the number of packets sent and periodicity of connections \cite{gu2008botminer,tegeler2012botfinder}. Other methods have used features specific to the TLS protocol to correlate the application's identity, server's identity, and behavior of the interaction to improve detection \cite{anderson16identifying}.

Website fingerprinting is another well studied encrypted traffic inference goal \cite{liberatore2006inferring, panchenko2016website, panchenko2011website, wang2014effective, wang2016realistically}. This problem is typically framed as the adversary attempting to identify connections to a small list of censored websites over the Tor network by leveraging side channel information such as the size of packet bursts and unique packet sizes \cite{wang2014effective}. Tor, especially when used with pluggable transports, makes website fingerprinting significantly more difficult, but the reliable detection of Tor pluggable transports has been demonstrated \cite{wang15seeing}.

More direct inferences on the body of encrypted HTTP messages have also been studied. One example of this class of attacks includes inferring the video a user is watching over popular streaming services \cite{reed2017identifying,schuster2017beauty}. Keyword fingerprinting is a recent attack that identifies individual queries sent to web applications, e.g., a search query to Google, over Tor \cite{oh2017fingerprinting}. Inferring the size of an encrypted resource is well known \cite{wagner1996analysis}, and has recently been used to identify users based on the unique, dynamic content the web server sends them \cite{van2016request}.

In contrast to previous work that makes inferences on the HTTP body, we introduce methods that infer the HTTP protocol semantics, thus understanding the protocol machinery of HTTP transactions inside of a TLS encrypted tunnel. Our results may provide value for many of the goals described in this section, either directly or indirectly as a means to normalize data features. 

\bgroup
\def\arraystretch{1.2}
\begin{table}[t!]\footnotesize
\center
\begin{tabular}{|l|r|r|r|}
\hline
\multirow{2}{*}{Dataset Name} & \multicolumn{1}{c|}{TLS}         & \multicolumn{1}{c|}{HTTP/1.1}    & \multicolumn{1}{c|}{HTTP/2}       \\
                              & \multicolumn{1}{c|}{Connections} & \multicolumn{1}{c|}{TX's} & \multicolumn{1}{c|}{TX's} \\
\hline
\hline
\texttt{firefox\_h} & 61,091 & 72,828 & 132,685 \\
\hline
\texttt{chrome\_h} & 379,734 & 515,022 & 561,666 \\
\hline
\texttt{tor\_h} & 6,067 & 50,799 & 0 \\
\hline
\texttt{malware\_h} & 86,083 & 182,498 & 14,734  \\
\hline
\texttt{enterprise\_m} & 171,542 & -- & --  \\
\hline
\texttt{malware\_m} & 73,936 & -- & --  \\
\hline
\texttt{tor\_open\_w} & 5,000 & 54,079 & 0 \\
\hline
\texttt{tor\_censor\_w} & 2,500 & 31,707 & 0 \\
\hline
\end{tabular}
\caption{Datasets ending with \_h are primarily used for the HTTP protocol semantics classification experiments (Section \ref{sec:header-inferences}), datasets ending with \_m are used for the malware classification experiments (Section \ref{sec:malware-detection}), and datasets ending with \_w are used for the website fingerprinting experiments (Section \ref{sec:website-fingerprinting}).}
\label{table:dataset_summary}
\end{table}
\egroup

\section{Data}
\label{sec:data}

\noindent Throughout this paper, we use a combination of data sources collected from automated runs in Linux-based virtual machines, a commercial malware analysis sandbox, and a real enterprise network. The Linux-based VMs used CentOS 7 running on VMware ESXi. The malware analysis sandbox executed tens-of-thousands of unique binaries each day under Windows 7 and 10, where we restricted our analysis to convicted samples that used TLS to communicate. The enterprise network had roughly 3,500 unique internal IP addresses per day. With the notable exception of the data collected from the enterprise network, we also collected the key material necessary to decrypt the connections in all datasets. This allowed us to correlate the decrypted HTTP transactions with the observable data features. A summary of the datasets is given in Table \ref{table:dataset_summary}, where datasets ending with \texttt{\_h} are used for the HTTP inference experiments, datasets ending with \texttt{\_m} are used for the malware detection experiments, and datasets ending with \texttt{\_w} are used for the website fingerprinting experiments.

\subsection{HTTP Inferences Data}
\label{sec:header-data}

\noindent To collect the ground truth for the application layer requests and responses, the Linux-based VMs contacted each site in the Alexa top-1,000 using Chrome 63.0, Firefox 58.0, and Tor Browser 7.0.11. This data collection was repeated each day for two weeks in December 2017. Two weeks of malware data were collected from a commercial malware analysis sandbox during October 2017. Varying approaches were taken to collect the key material necessary to decrypt the connections for each dataset as explained in each subsection. When a given network connection failed or decryption of that connection failed, we discarded the sample resulting in different datasets sizes. Session decryption failures occurred due to occasional key extraction problems, e.g., the encryption keys were not in memory during the memory snapshots for Tor. This occurred uniformly at random and thus unlikely to introduce bias.

The Firefox 58.0, Chrome 63.0, and Tor Browser 7.0.11 datasets used in this section are in the process of being open sourced. The dataset contains packet captures of the TLS/Tor sessions and the encryption keys needed to decrypt the TLS/Tor sessions for the \texttt{firefox\_h}, \texttt{chrome\_h}, and \texttt{tor\_h} datasets.

\subsubsection{Firefox 58.0 and Chrome 63.0}

\noindent Both Firefox 58.0 and Chrome 63.0 support the export of TLS 1.0-1.3 master secrets through the \texttt{SSLKEYLOGFILE} environment variable. To prepare data collection for a given browser and site pair, we first set the \texttt{SSLKEYLOGFILE} environment variable and begin collecting the network traffic with \texttt{tcpdump}. Then we launch the specified browser and site pair in private mode using the \texttt{Xvfb} virtual window environment, and we allow the process to run for 15 seconds. After 15 seconds, all associated processes are killed, and we store the packet capture and TLS master secrets for additional processing as described in Section \ref{sec:decrypting}.

For Firefox, we decrypted a total of 31,175 HTTP/1.1 and 29,916 HTTP/2 connections. For Chrome, we decrypted a total of 242,036 HTTP/1.1 and 137,698 HTTP/2 connections. We omitted browser-dependent connections from our results, e.g., pocket recommendations in Firefox.

\subsubsection{Tor Browser 7.0.11}
\label{sec:tor-browser}

\noindent A similar structure to the Firefox/Chrome data collection is followed for Tor Browser, except that Tor Browser 7.0.11 explicitly prevents the export of its key material due to security concerns. For this reason, instead of setting the environment variable, we take memory snapshots of the \texttt{tor} and \texttt{firefox} processes every 3 seconds after the first second. The information in \texttt{/proc/<pid>/maps} and \texttt{/proc/<pid>/mem} is used to associate the correct memory to the process ids. These memory dumps are then post-processed as described in Section \ref{sec:extracting_keys} to extract the needed key material.

We decrypted a total of 6,067 TLS-Tor connections and 50,799 HTTP/1.1 transactions. If we failed to decrypt the Tor tunnel or one of the underlying streams, the sample was discarded. The difference in the number of connections between the Tor dataset and the Firefox/Chrome datasets was due to Tor multiplexing many unique streams over a single connection.

\subsubsection{Malware Sandbox}
\label{sec:malware-sandbox}

\noindent Each convicted malware sample was executed in either a Windows 7 or Windows 10 virtual machine for 5 minutes. After the 5 minute window, the packet capture was stored and the virtual machine's full memory snapshot was analyzed as described in Section \ref{sec:extracting_keys}. Man-in-the-middle or other more intrusive means to collect the key material was decided against to avoid contaminating the behavior of the malware. This decision did result in fewer decrypted TLS connections than total TLS connections because the TLS library could zeroize the memory containing the key material. That being said, we were still able to decrypt $\sim$80\% of the TLS connections.

For the malware dataset, we decrypted a total of 82,177 HTTP/1.1 and 3,906 HTTP/2 connections. We omitted browser-dependent and VM-dependent connections from our results, e.g., connections to \texttt{ieonline.microsoft.com}. We did not perform any other filtering besides these obvious filters, i.e., we did not attempt to distinguish between legitimate malicious traffic and benign CDN connections.

This dataset is significantly more heterogeneous than the previous datasets due to the malware samples not being restricted to a single TLS library or a prescribed set of websites. $\sim$70\% of the malware samples used the \texttt{Schannel} library provided by the operating system, with the remaining samples using a variety of alternatives.

\subsubsection{Extracting the Key Material}
\label{sec:extracting_keys}

\noindent To decrypt the packet capture from a convicted malware sample (Section \ref{sec:malware-sandbox}) or a Tor instance (Section \ref{sec:tor-browser}), we extracted key material from the memory snapshot taken in the final moments of the malware's execution or from a series of snapshots during the lifetime of the Tor process. Prior work on scanning RAM for key material exists \cite{halderman09:coldboot,kambic16:schannel,ligh14:artofmem}, but prior techniques were neither sufficiently lightweight nor general enough to directly integrate into the commercial malware analysis sandbox. The production use case required fully automated forensic analysis of a previously unknown executable, under strict CPU and time constraints.  Our approach instead leveraged the fact that malware primarily uses established TLS libraries, especially ones built into the targeted platform \cite{anderson16deciphering}.

The cornerstone of key extraction is the fact that TLS libraries tend to nest the master secret in a predictable data structure, e.g., for OpenSSL:
\lstset{
  basicstyle=\footnotesize\ttfamily,
  breaklines=true,
  escapeinside=||
}
\begin{Ccode}
struct ssl_session_st {
  int ssl_version;
  unsigned int key_arg_length;
  unsigned char key_arg[8];
  int master_key_length; // 48
  unsigned char master_key[48];
  unsigned int session_id_length;
  unsigned char session_id[32];
  ...
\end{Ccode}
In memory, this data structure appears as:
\lstset{
  basicstyle=\footnotesize\ttfamily,
  breaklines=true,
  escapeinside=||
}
\begin{lstlisting}
  |\white{03 03 00 00 00 00 00 00 00 00 00 00 00 00 00 00}|
  |\white{30 00 00 00}\hspace{0.8mm}\hilite{44 0E 70 5C 1C 22 45 07 6C 1C ED 0D}|
  |\hilite{E3 74 DF E2 C9 71 AF 41 2C 0B E6 AF 70 32 6E C3}|
  |\hilite{A3 2C A0 E6 3A 7A FF 0E F3 70 A2 8A 88 52 B2 2D}|
  |\hilite{D1 B3 F6 F2}\hspace{0.8mm}\white{20 00 00 00 CD 31 58 BF DF 97 B0 F8}|
  |\white{C0 86 BA 48 47 93 B0 A5 BA C1 5B 4B 35 37 7F 98}|
\end{lstlisting}
where the leading \texttt{0x0303} indicates TLS 1.2, and the 48-byte master secret is highlighted.  It was therefore straightforward to write a regular expression that yielded all OpenSSL master secrets in memory within seconds (BoringSSL and Microsoft Schannel were similar). Mozilla NSS allocated the TLS master secret as a standalone 48-byte buffer, which could in principle be anywhere on the heap with no guaranteed context. However, we discovered that in practice NSS consistently allocated the TLS master secret directly adjacent to a predictable data structure: the struct that carried the pointer to the master secret. This held true across multiple operating systems and platforms, and we were able to reliably extract NSS master secrets using regular expressions. We were able to extract the Tor 128-bit AES keys in a similar manner. We used the following regular expressions to extract TLS master secrets and AES keys:
{\small
\begin{align}
\texttt{BoringSSL}:\hspace{1mm} & \texttt{(\hex{02}\hex{00}|[\hex{00}-\hex{03}]\hex{03})\hex{00}\hex{00}(?=} \nonumber \\
                                & \hspace{0mm} \texttt{.\{2\}.\{2\}\hex{30}\hex{00}\hex{00}\hex{00}(.\{48\})[\hex{00}-} \nonumber \\
                                & \hspace{0mm} \texttt{\hex{20}]\hex{00}\hex{00}\hex{00})}\nonumber \\
\texttt{NSS}:\hspace{1mm}       & \texttt{\hex{11}\hex{00}\hex{00}\hex{00}(?=(.\{8\}\hex{30}\hex{00}\hex{00}} \nonumber \\
                                & \hspace{0mm}\texttt{\hex{00}|.\{4\}.\{8\}\hex{30}\hex{00}\hex{00}\hex{00}.\{4\})} \nonumber \\
                                & \hspace{0mm} \texttt{(.\{48\}))} \nonumber \\
\texttt{OpenSSL}:\hspace{1mm}   & \texttt{(\hex{02}\hex{00}|[\hex{00}-\hex{03}]\hex{03})\hex{00}\hex{00}(?=} \nonumber \\
                                & \hspace{0mm}\texttt{.\{4\}.\{8\}\hex{30}\hex{00}\hex{00}\hex{00}(.\{48\})[\hex{00}-}\nonumber \\
                                & \hspace{0mm}\texttt{\hex{20}]\hex{00}\hex{00}\hex{00})} \nonumber \\
\texttt{Schannel}:\hspace{1mm}  & \texttt{\hex{35}\hex{6c}\hex{73}\hex{73}(?=(\hex{02}\hex{00}|[\hex{00}-} \nonumber \\
                                & \hspace{0mm}\texttt{\hex{03}]\hex{03})\hex{00}\hex{00}(.\{4\}.\{8\}.\{4\})} \nonumber \\
                                & \hspace{0mm} \texttt{(.\{48\}))} \nonumber \\
\texttt{Tor-AES}:\hspace{1mm}   & \texttt{\hex{11}\hex{01}\hex{00}\hex{00}\hex{00}\hex{00}\hex{00}\hex{00}(?=} \nonumber \\
                                & \hspace{0mm}\texttt{(.\{16\})(.\{16\}))} \nonumber \\ \nonumber
\end{align}
}
\vspace{-11mm}

\subsubsection{Decrypting the sessions}
\label{sec:decrypting}

\begin{figure}[t!]
\begin{center}
\lstset{
    string=[s]{"}{"},
    stringstyle=\color{blue},
    comment=[l]{:},
    commentstyle=\color{black},
}
\begin{lstlisting}[
%linewidth=7.5cm,
basicstyle=\small\ttfamily,
]
{
 "tls_records": [
   ...
   {
     "type": "app_data",
     "length": 1052,
     "decrypted_data": {
       "protocol": "Tor",
       "length": 1028,
       "cells": [
         {
           "circ_id": "xxxxxxxx",
           "cell_type": "RELAY",
           "command": "RELAY_DATA",
           "stream_id": "xxxx",
           "digest": "xxxxxxxx",
           "length": 340,
           "decrypted_data": {
             "tls_records": [
               {
                 "type": "app_data",
                 "length": 335,
                 "decrypted_data": {
                   "method": "GET",
                   "uri": "/",
                   "v": "HTTP/1.1",
                   "headers": [
                     ...
                   ],
             ...
\end{lstlisting}
\end{center}
\caption{An example of a decrypted HTTP/1.1 GET request tunneled over Tor, represented in JSON.}
\label{fig:tor-json}
\end{figure}

\noindent We wrote a custom tool to decrypt a packet capture given a file containing the extracted key material. Our tool supports the decryption of SSL 2.0-TLS 1.3, 200+ cipher suites, and can parse the HTTP/1.x, HTTP/2, and Tor application layer protocols. For Tor traffic, it can also decrypt the underlying \texttt{RELAY} and \texttt{RELAY\_EARLY} cells. If a decrypted stream contains a TLS session, the stream will in turn be decrypted and the resulting application layer data will be extracted. The results of the decryption program are stored in JSON for convenient manipulation by the machine learning preprocessors. As an example, Figure \ref{fig:tor-json} illustrates the decrypted output of an HTTP/1.1 \texttt{GET} request tunneled over Tor.

Browsers that support exporting the TLS key material through the \texttt{SSLKEYLOGFILE} environment variable adhere to the NSS key log format, which associates the TLS \texttt{client\_random} to a TLS master secret. The Tor and malware datasets do not support this functionality, and we were forced to create a new format that omitted the \texttt{client\_random}. In this case, we brute force the decryption by attempting to decrypt a connection with all of the extracted keys. For TLS 1.2, this involves decrypting the small \texttt{finished} message, which is relatively efficient. We attempt all master secrets until the message is properly decoded. For the Tor \texttt{RELAY} cells, we again try all available AES keys, making sure to not disrupt the state of the previous onion layer's stream cipher in the case of a decryption failure. Once we properly decrypt the \texttt{RELAY} cell by identifying a valid relay command and recognized field, we add the cipher to the circuit's cipher list and return the decrypted data.

\subsection{Malware Classification}
\label{section:malware-data}

\noindent For our malware classification experiments, we used malware data from the same malware analysis sandbox as above collected in November and December, 2017. The enterprise network data was also collected during the same time frame, but was uniformly sampled to avoid severe class imbalance.

We collected the packet captures for each malware run, but ignored the key material. We processed each packet capture to extract the encrypted data in a similar format to the previous section, but did not use the decryption feature. We were able to associate the hash of the process initiating the TLS connection with the 5-tuple of the connection, and discarded any connections that were not initiated by executables that were flagged as malware. We further cleaned the dataset by discarding popular CDNs and analytics sites such as \texttt{gstatic.com} and \texttt{google-analytics.com}. This may have been overly aggressive, but after manually inspecting the decrypted contents of several hundred of these samples, we concluded that they are much more likely to be benign. Finally, the number of unique TLS \texttt{server\_name} values was also kept to a maximum of 50 uniformly at random samples per month to avoid skewing our results towards the performance on popular domains. Post-filtering, we were left with 34,872 and 39,064 malicious TLS connections for November and December, respectively.

For the benign dataset, we collected TLS connections from a real enterprise network using our tools during November and December, 2017. We did not have access to any key material in this case, and obviously did not perform decryption. We filtered this data with freely available IP blacklists. For the reasons described above, we only allowed a maximum of 50 unique TLS \texttt{server\_name} values per month, chosen uniformly at random, in the enterprise data. The mean number of unique \texttt{server\_name} values per month was $\sim$5, which was increased by an order of magnitude to maintain some information about prevalence. After uniformly sampling and filtering, we were left with 87,016 and 84,526 benign TLS connections for November and December, respectively.

\subsection{Website Fingerprinting}

\noindent We aimed to emulate the standard website fingerprinting open world experiment \cite{panchenko2016website, wang2016realistically}. The data was collected in a similar fashion to what was described in Section \ref{sec:tor-browser}, but with different website lists. While we did extract the key material, we did not use the decrypted data to train the website fingerprinting algorithms.

We used Tor Browser 7.0.11 to connect to each site in a list of 50 censored websites. We repeated this cycle until we were able to collect data from 50 successful connections that were able to be decrypted for each censored website. This data collection was performed during the second week of January 2017.

During the second week of January 2017, we also used our Tor Browser data collection strategy while connecting to each site in the Alexa top-10k. We excluded any samples that failed to decrypt and sites that appeared in the list of monitored sites. Similar to previous work, we took the top 5,000 sites that remained.

\section{Inferring HTTP Protocol Semantics}
\label{sec:header-inferences}

\noindent Our framework to infer various attributes of HTTP in encrypted network connections relies heavily on the labeled data of Section \ref{sec:data}. Given that data, it is possible to make a number of interesting inferences on encrypted HTTP transactions without having to perform decryption. We used a standard random forest classifier with 100 trees for all experiments because they have been shown to be a superior choice for network traffic analysis tasks \cite{anderson2017noisy}. While alternative machine learning methods could prove to be more performant, these investigations are neither the focus nor in scope for this paper.

We report both the raw accuracy and the unweighted $F_1$ score for each problem. As explained in Section \ref{sec:inferred-http-headers}, several of the inference problems are posed as multi-class classification, and the unweighted $F_1$ score provides a better representation of the classifier's performance on the minority classes. It is defined as the unweighted mean of the $F_1(L_i)$ scores for each label $L_i$ in the multi-class problem, where $F_1(L_i)$ is defined as:
\begin{equation}
\text{$F_1$($L_i$)} = 2 \times \frac{\text{precision}_i \times \text{recall}_i}{\text{precision}_i + \text{recall}_i}
\end{equation}

For all results, we segment the two weeks of data described in Section \ref{sec:header-data} into training and testing datasets. The first week of data is used for training, and the second week of data is used for testing. Table \ref{table:summary_header_results} provides a summary of all inference results.

\subsection{Data Features}
\label{sec:http-header-data-features}

\noindent We use two categories of data features to classify HTTP protocol semantics: features that are dependent on the location (relative to the surrounding TLS records) of the target TLS record containing the HTTP request or response, and features derived from all packets in a connection. For the location-specific feature set, we analyze the current, preceding 5, and following 5 TLS records. For each TLS record, we extract:
\begin{enumerate}[itemsep=-1mm]
\item The number of packets
\item The number of packets with the TCP \texttt{PUSH} flag set
\item The average packet size in bytes
\item The type code of the TLS record
\item The TLS record size in bytes
\item The direction of the TLS record
\end{enumerate} 
We treat the counts and sizes as real-valued features, the TLS type code as a categorical feature, and the direction as a categorical feature where 0 indicates client $\rightarrow$ server, 1 indicates server $\rightarrow$ client, and 2 indicates no TLS record. All features except direction are set to 0 if a TLS record does not exist, e.g., features related to the following 5 TLS records when the target TLS record ends the connection. We ignored timing-based features because we found them to be unreliable.

For the connection-dependent features, we extracted the number of packets, number of packets with the TCP \texttt{PUSH} flag set, and the average packet size separately for each direction of the connection. We also extracted the sizes in bytes of the first 100 TLS records, where the size is defined to be negative if the record was sent by the server. This array was null padded. Finally, we computed the connection's total duration in seconds. All of these values were represented as real-valued features.

Each sample for the classification problems discussed in this section is composed of 174 data features: 66 record-dependent features, 6 features extracted from each of the 11 TLS records analyzed, and 108 connection-dependent features. The Tor experiments are the exception because we omit the connection-dependent features. We found these features to be unreliable in the Tor HTTP protocol semantics inference task, which is not surprising considering the number of unique TLS connections multiplexed over a single Tor tunnel.

\subsection{Inferred HTTP Protocol Semantics}
\label{sec:inferred-http-headers}

\noindent Before we can infer the values contained within an HTTP request or response, we need to be able to identify which TLS records contain a request or response. In our results, this problem is labeled ``message-type", and it is a binary classification problem where the labels indicate if a TLS record contains at least one HTTP request or response. We chose this approach because it lets us ignore many of the complexities associated with HTTP/2 frame types and Tor cell types.

For HTTP requests, we study 2 multi-class classification problems: the \texttt{method} and \texttt{Content-Type} fields, and 3 binary classification problems: the \texttt{Cookie}, \texttt{Referer}, and \texttt{Origin} fields. For the binary classification problems, we are attempting to determine if the field key appears one or more times in the HTTP request.

For HTTP responses, we study 3 multi-class classification problems: the \texttt{status-code}, \texttt{Content-Type}, and \texttt{Server} fields, and 4 binary classification problems: the \texttt{Access-Control-Allow-Origin}, \texttt{Via}, \texttt{Accept-Ranges}, and \texttt{Set-Cookie} fields.

We focused on this set of problems because they were well-represented in both our HTTP/1.1 and HTTP/2 datasets and they exhibited a reasonable level of diversity. As one would expect given our data collection strategy, our problem selection is biased towards HTTP response fields. As explained in Section \ref{sec:limitations-and-future-work}, we believe the approach outlined in this paper would translate to a larger set of request-related problems if appropriate training data was available.

\subsubsection{Multi-Class Labels}
\label{sec:inferred-http-headers-labels}

\noindent Table \ref{table:multi-class-header-labels} lists the labels for all multi-class classification problems. There are some instances of ambiguity in the HTTP request and response field values. For example, the ``application/octet" value for the response \texttt{Content-Type} field can be used for multiple file types, and the ``nginx" value for the \texttt{Server} field can map to multiple version. For our experiments, we take the field value as is and do not attempt to relabel samples.


\bgroup
\def\arraystretch{1.2}
\begin{table}[t!]\small
\center
\begin{tabular}{|l|l|l|}
\hline
\multicolumn{1}{|c|}{Problem}       & \multicolumn{1}{c|}{HTTP/1.1} & \multicolumn{1}{c|}{HTTP/2} \\
              & \multicolumn{1}{c|}{Label Set} & \multicolumn{1}{c|}{Label Set} \\
\hline
\hline
method        & GET, POST, & GET, POST, \\
(req)         & OPTIONS, HEAD, & OPTIONS, HEAD \\
              & PUT  &  \\
\hline
Content       & json, plain & json, plain \\
-Type         &  &  \\
(req)         &  &  \\
\hline
\hline
status        & 100, 200, 204, 206, & 200, 204, 206, \\
-code         & 302, 303, 301, & 301, 302, 303, \\
(resp)        & 304, 307, 404 & 304, 307, 404 \\
              &  &  \\
\hline
Content       & html, javascript, & html, javascript, \\
-type         & image, video, css, & image, video, css, \\
(resp)        & octet, json, font, & octet, son, font, \\
              & plain & plain, protobuf \\
              &  &  \\
\hline
Server        & nginx-1.13/1.12, & nginx-1.13/1.12, \\
(resp)        & nginx-1.11/1.10/1.8, & nginx-1.11/1.10/1.6, \\
              & nginx-1.7/1.4, nginx, & nginx-1.4/1.3, nginx, \\
              & cloudflare-nginx, & cloudflare-nginx, \\
              & openresty, Apache, & Apache, Coyote/1.1, \\
              & Coyote/1.1, & IIS/8.5, Golfe2, sffe, \\
              & AmazonS3, & cafe, ESF, GSE, \\
              & NetDNA/2.2, & gws, UploadServer, \\
              & IIS-7.5/8.5, & Akamai, Google, \\
              & jetty-9.4/9.0 & Dreamlab, Tengine, \\
              &  & AmazonS3, \\
              &  & NetDNA/2.2 \\
\hline
\end{tabular}
\caption{Label sets for the multi-class HTTP protocol semantics inference experiments. For HTTP/1.1, there are 5 \texttt{method}, 2 request \texttt{Content-Type}, 10 \texttt{status-code}, 9 response \texttt{Content-Type}, and 18 \texttt{Server} labels. For HTTP/2, there are 4 \texttt{method}, 2 request \texttt{Content-Type}, 9 \texttt{status-code}, 10 response \texttt{Content-Type}, and 25 \texttt{Server} labels.}
\label{table:multi-class-header-labels}
\end{table}
\egroup

\begin{algorithm}[t!]\small
\caption{Iterative HTTP Protocol Semantics Inference}
\begin{algorithmic}[1]
\Procedure{iterative\_semantics\_classify}{}
  \given:
    \State $\textit{conn} := \text{features describing connection}$
  \State $\textit{alp} \gets \texttt{application\_layer\_protocol}(\textit{conn})$
  \State $\textit{recs} \gets \texttt{classify\_message\_types}(\textit{conn}, \textit{alp})$
  \For {$\textit{rec} \in \textit{recs}$}:
    \If {$\textit{rec.type} \neq \texttt{Headers}$}:
      \continue
    \EndIf
    \State $\texttt{get\_record\_features}(\textit{rec}, \textit{alp})$
    \State $\texttt{classify\_semantics}(\textit{rec}, \textit{alp})$
  \EndFor
  \While {not converged}:
    \For {$\textit{rec} \in \textit{recs}$}:
      \If {$\textit{rec.type} \neq \texttt{Headers}$}:
        \continue
      \EndIf
      \State $\texttt{get\_record\_features}(\textit{rec}, \textit{alp})$
      \State $\texttt{get\_inferred\_features}(\textit{rec}, \textit{alp})$
      \State $\texttt{classify\_semantics}(\textit{rec}, \textit{alp})$
    \EndFor  
  \EndWhile
\EndProcedure
\end{algorithmic}
\label{alg:iterative}
\end{algorithm}

\subsection{Iterative Classification}
\label{sec:iterative-classification}

\begin{figure*}[t!]
\centering
\subfloat[chrome\_h] {
	\hspace{.02cm}
   	\includegraphics[scale=0.063,valign=t]{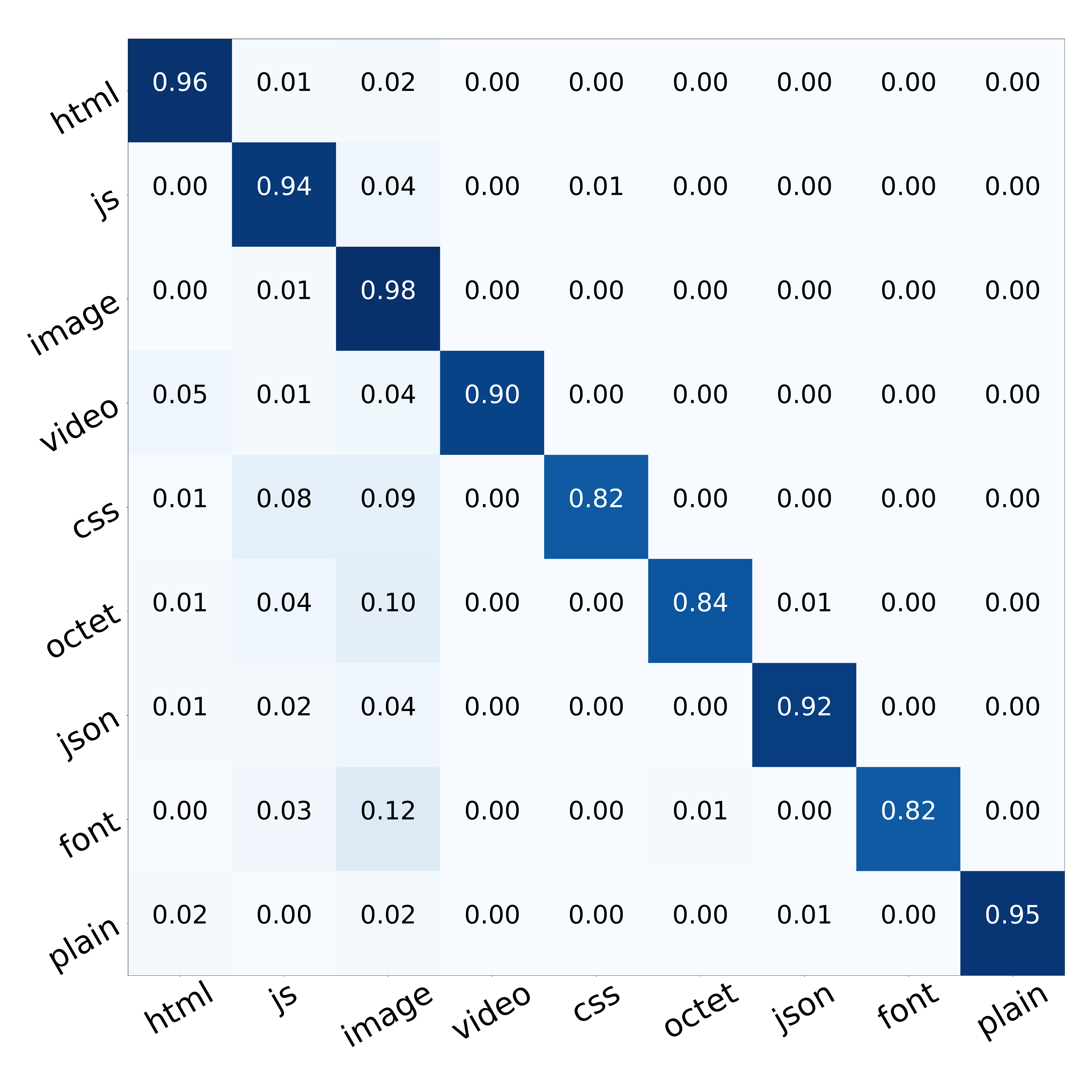}
	\label{fig:confusion-http-chrome-content-type-primary}
}
\hspace{-3.75mm}
\subfloat[malware\_h] {
   	\includegraphics[scale=0.063,valign=t]{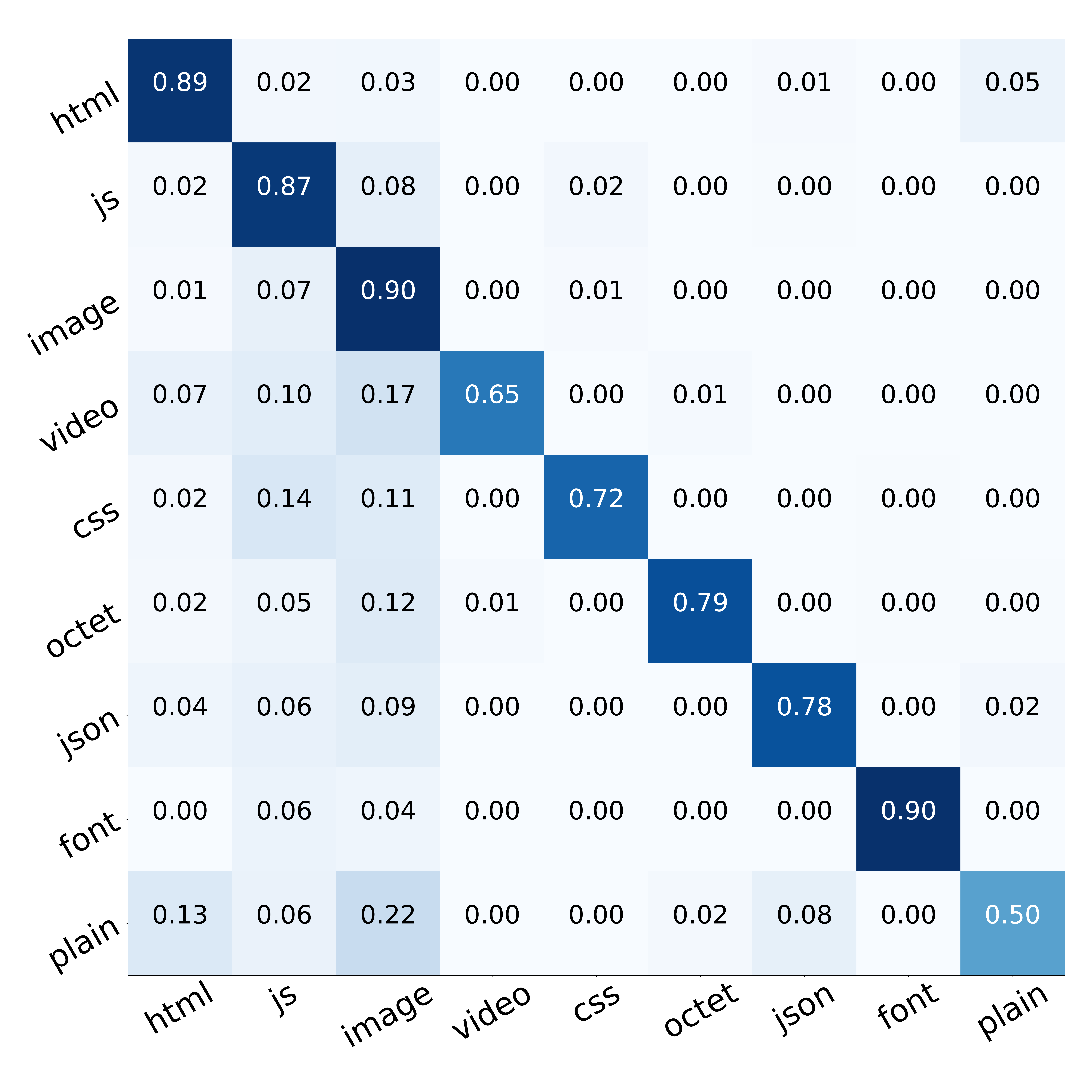}
	\label{fig:confusion-http-malware-content-type-primary}
}
\hspace{-3.75mm}
\subfloat[tor\_h] {
   	\includegraphics[scale=0.063,valign=t]{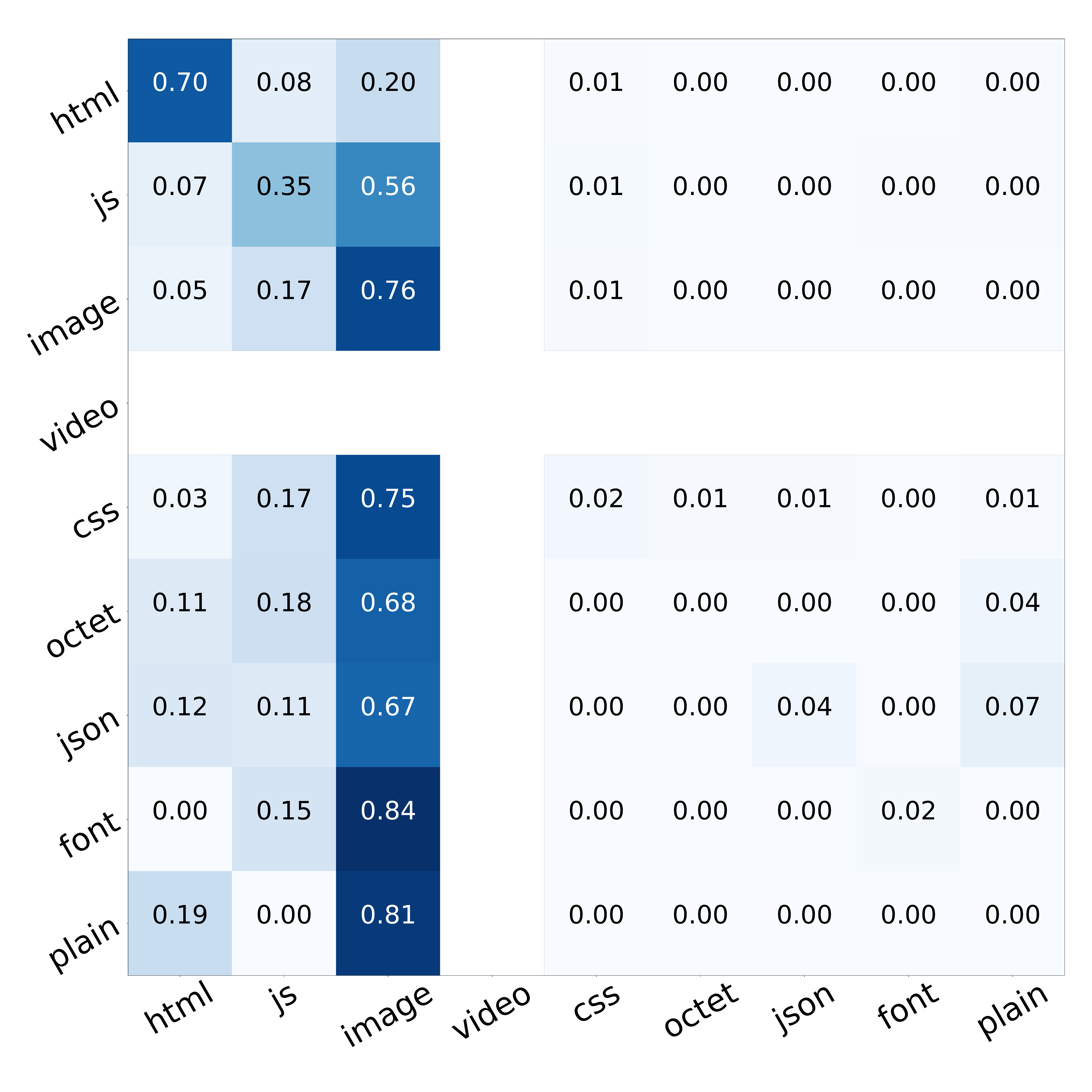}
	\label{fig:confusion-http-tor-content-type-primary}
}
\caption{Confusion matrices for the HTTP/1.1 response \texttt{Content-Type} header field value on the chrome\_h, malware\_h, and tor\_h datasets. Matrix elements are left blank if there were no relevant labels in a particular dataset.}
\label{fig:http-confusion-matrices-primary}
\end{figure*}

\noindent Many of the inference goals outlined in the previous section are dependent on each other, e.g., the value of the response \texttt{Content-Type} is correlated with the \texttt{Server} value, or a response \texttt{Content-Type} is correlated with other response \texttt{Content-Type}'s in the same TLS connection. We take this fact into account by using an iterative classification framework.

Given an encrypted TLS connection, we first determine the application layer protocol (\texttt{alp}) through the TLS \texttt{application\_layer\_protocol\_negotiation} extension. If this extension is absent, we use a classifier based on the first 20 TLS record lengths to classify the connection as either HTTP/1.1 or HTTP/2. Given \texttt{alp} and the data features described in Section \ref{sec:http-header-data-features}, we use a binary classifier to identify each TLS \texttt{application\_data} record containing HTTP header fields. In this section (but not Section \ref{sec:use_cases}), we discard connections with inaccurately labeled TLS records, e.g., we classify an HTTP/2 \texttt{HEADERS} frame as a \texttt{DATA} frame. Although this process resulted in a $<$1\% reduction in the total number of samples classified, this process is important to note while interpreting this section's results.

For each TLS record identified as containing HTTP header fields, we extract the Section \ref{sec:http-header-data-features} data features and then apply the classifiers related to the request semantics for client records and the response semantics for server records. At this point, we associate the original record's data features with the classifier outputs for all records containing HTTP header fields in the same connection excluding the target classification problem in the target record. This enhanced feature set has length 68 for HTTP/1.1 and length 74 for HTTP/2. The subcomponents of the enhanced feature vector correspond to the sum of all other predicted outputs from the previous iteration after the predicted outputs have been translated to an indicator vector. For example, if there is a connection with 7 HTTP requests and the \texttt{Referer} field was present in 4 out of the 6 non-target requests, then the subcomponent of the enhanced feature vector related to the \texttt{Referer} field would be $[2,4]$. Given the enhanced features, the HTTP protocol semantics are classified using the TLS record's data features and the inferences from the previous iteration. We consider the algorithm converged when no predicted outputs change value. In our experiments, the iterative algorithm typically converged in 2 and at most 4 iterations.

Algorithm \ref{alg:iterative} summarizes the iterative classification procedure. It uses multiple, intermediate classifiers, each of which needs to be trained. These infer the application layer protocol, the TLS records that contain the protocol semantics of HTTP requests and responses, the HTTP protocol semantics given only the target record's features, and the HTTP protocol semantics given all features. When classifying an unseen test sample, two classifiers per inference will be needed to carry out Algorithm \ref{alg:iterative}.

Tor necessitates a minor exception to the iterative algorithm due to the large number of unique TLS connections that are multiplexed over a single TLS/Tor connection. Instead of using all inferred HTTP values within a connection for the enhanced feature vector, we only use the predicted outputs of the preceding and following 5 HTTP transactions in the Tor connection.

\bgroup
\def\arraystretch{1.08}
\begin{table*}\small
\center
\begin{tabular}{|l|l||c|c||c|c|||c|c||c|c|}
\hline
\multirow{3}{*}{Problem} & \multirow{3}{*}{Dataset} & \multicolumn{4}{c|||}{HTTP/1.1} & \multicolumn{4}{c|}{HTTP/2} \\
\cline{3-10}
                                                   & & \multicolumn{2}{c||}{Single Pass} & \multicolumn{2}{c|||}{Iterative}                                                      & \multicolumn{2}{c||}{Single Pass} & \multicolumn{2}{c|}{Iterative} \\
\cline{3-10}
                                                     & & $F_1$ Score & Acc & $F_1$ Score & Acc & $F_1$ Score & Acc & $F_1$ Score & Acc \\
\hline
\hline

\multirow{ 4}{*}{\parbox{3cm}{\texttt{message-type}}} & firefox\_h & \cellcolor{green!97!yellow!35!white}0.996 & \cellcolor{green!97!yellow!35!white}0.996 & & & \cellcolor{green!92!yellow!35!white}0.987 & \cellcolor{green!92!yellow!35!white}0.991 & & \\ 
& chrome\_h & \cellcolor{green!94!yellow!35!white}0.991 & \cellcolor{green!94!yellow!35!white}0.993 & & & \cellcolor{green!90!yellow!35!white}0.986 & \cellcolor{green!90!yellow!35!white}0.986 & & \\ 
& malware\_h & \cellcolor{green!97!yellow!35!white}0.995 & \cellcolor{green!97!yellow!35!white}0.996 & & & \cellcolor{green!90!yellow!35!white}0.981 & \cellcolor{green!90!yellow!35!white}0.989 & & \\ 
& tor\_h & \cellcolor{green!15!yellow!35!white}0.869 & \cellcolor{green!15!yellow!35!white}0.878 & & & & & & \\ 
\hline
\hline
\multirow{ 4}{*}{\parbox{3cm}{\texttt{method}}} & firefox\_h & \cellcolor{green!66!yellow!35!white}0.909 & \cellcolor{green!66!yellow!35!white}0.990 & \cellcolor{green!79!yellow!35!white}0.943 & \cellcolor{green!79!yellow!35!white}0.995 & \cellcolor{green!35!yellow!35!white}0.815 & \cellcolor{green!35!yellow!35!white}0.992 & \cellcolor{green!95!yellow!35!white}0.989 & \cellcolor{green!95!yellow!35!white}0.997 \\ 
& chrome\_h & \cellcolor{green!87!yellow!35!white}0.968 & \cellcolor{green!87!yellow!35!white}0.995 & \cellcolor{green!92!yellow!35!white}0.978 & \cellcolor{green!92!yellow!35!white}0.998 & \cellcolor{green!60!yellow!35!white}0.888 & \cellcolor{green!60!yellow!35!white}0.994 & \cellcolor{green!78!yellow!35!white}0.936 & \cellcolor{green!78!yellow!35!white}0.999 \\ 
& malware\_h & \cellcolor{yellow!98!red!35!white}0.701 & \cellcolor{yellow!98!red!35!white}0.994 & \cellcolor{green!0!yellow!35!white}0.705 & \cellcolor{green!0!yellow!35!white}0.996 & \cellcolor{yellow!94!red!35!white}0.699 & \cellcolor{yellow!94!red!35!white}0.985 & \cellcolor{yellow!90!red!35!white}0.687 & \cellcolor{yellow!90!red!35!white}0.985 \\ 
& tor\_h & \cellcolor{yellow!68!red!35!white}0.678 & \cellcolor{yellow!68!red!35!white}0.927 & \cellcolor{green!36!yellow!35!white}0.846 & \cellcolor{green!36!yellow!35!white}0.965 & & & & \\ 
\hline
\multirow{ 4}{*}{\parbox{3cm}{\texttt{Content-Type}}} & firefox\_h & \cellcolor{green!85!yellow!35!white}0.973 & \cellcolor{green!85!yellow!35!white}0.982 & \cellcolor{green!81!yellow!35!white}0.967 & \cellcolor{green!81!yellow!35!white}0.978 & \cellcolor{green!40!yellow!35!white}0.905 & \cellcolor{green!40!yellow!35!white}0.917 & \cellcolor{green!89!yellow!35!white}0.982 & \cellcolor{green!89!yellow!35!white}0.985 \\ 
& chrome\_h & \cellcolor{green!80!yellow!35!white}0.962 & \cellcolor{green!80!yellow!35!white}0.979 & \cellcolor{green!90!yellow!35!white}0.977 & \cellcolor{green!90!yellow!35!white}0.993 & \cellcolor{green!82!yellow!35!white}0.970 & \cellcolor{green!82!yellow!35!white}0.979 & \cellcolor{green!98!yellow!35!white}0.998 & \cellcolor{green!98!yellow!35!white}0.998 \\ 
& malware\_h & \cellcolor{yellow!73!red!35!white}0.796 & \cellcolor{yellow!73!red!35!white}0.825 & \cellcolor{green!29!yellow!35!white}0.888 & \cellcolor{green!29!yellow!35!white}0.900 & \cellcolor{yellow!4!red!35!white}0.624 & \cellcolor{yellow!4!red!35!white}0.788 & \cellcolor{yellow!65!red!35!white}0.711 & \cellcolor{yellow!65!red!35!white}0.887 \\ 
& tor\_h & \cellcolor{yellow!0!red!35!white}0.572 & \cellcolor{yellow!0!red!35!white}0.781 & \cellcolor{green!13!yellow!35!white}0.836 & \cellcolor{green!13!yellow!35!white}0.904 & & & & \\ 
\hline
\multirow{ 4}{*}{\parbox{3cm}{\texttt{Cookie} (b)}} & firefox\_h & \cellcolor{green!73!yellow!35!white}0.954 & \cellcolor{green!73!yellow!35!white}0.965 & \cellcolor{green!80!yellow!35!white}0.967 & \cellcolor{green!80!yellow!35!white}0.974 & \cellcolor{green!15!yellow!35!white}0.864 & \cellcolor{green!15!yellow!35!white}0.882 & \cellcolor{green!62!yellow!35!white}0.941 & \cellcolor{green!62!yellow!35!white}0.948 \\ 
& chrome\_h & \cellcolor{green!80!yellow!35!white}0.970 & \cellcolor{green!80!yellow!35!white}0.970 & \cellcolor{green!84!yellow!35!white}0.977 & \cellcolor{green!84!yellow!35!white}0.977 & \cellcolor{green!42!yellow!35!white}0.909 & \cellcolor{green!42!yellow!35!white}0.918 & \cellcolor{green!70!yellow!35!white}0.953 & \cellcolor{green!70!yellow!35!white}0.958 \\ 
& malware\_h & \cellcolor{green!34!yellow!35!white}0.900 & \cellcolor{green!34!yellow!35!white}0.902 & \cellcolor{green!44!yellow!35!white}0.916 & \cellcolor{green!44!yellow!35!white}0.918 & \cellcolor{green!0!yellow!35!white}0.837 & \cellcolor{green!0!yellow!35!white}0.864 & \cellcolor{green!36!yellow!35!white}0.898 & \cellcolor{green!36!yellow!35!white}0.913 \\ 
& tor\_h & \cellcolor{yellow!47!red!35!white}0.734 & \cellcolor{yellow!47!red!35!white}0.809 & \cellcolor{yellow!59!red!35!white}0.756 & \cellcolor{yellow!59!red!35!white}0.823 & & & & \\ 
\hline
\multirow{ 4}{*}{\parbox{3cm}{\texttt{Referer} (b)}} & firefox\_h & \cellcolor{green!76!yellow!35!white}0.948 & \cellcolor{green!76!yellow!35!white}0.981 & \cellcolor{green!86!yellow!35!white}0.969 & \cellcolor{green!86!yellow!35!white}0.989 & \cellcolor{green!70!yellow!35!white}0.930 & \cellcolor{green!70!yellow!35!white}0.982 & \cellcolor{green!78!yellow!35!white}0.950 & \cellcolor{green!78!yellow!35!white}0.987 \\ 
& chrome\_h & \cellcolor{green!86!yellow!35!white}0.968 & \cellcolor{green!86!yellow!35!white}0.993 & \cellcolor{green!90!yellow!35!white}0.978 & \cellcolor{green!90!yellow!35!white}0.995 & \cellcolor{green!59!yellow!35!white}0.892 & \cellcolor{green!59!yellow!35!white}0.986 & \cellcolor{green!74!yellow!35!white}0.933 & \cellcolor{green!74!yellow!35!white}0.991 \\ 
& malware\_h & \cellcolor{green!45!yellow!35!white}0.914 & \cellcolor{green!45!yellow!35!white}0.923 & \cellcolor{green!54!yellow!35!white}0.928 & \cellcolor{green!54!yellow!35!white}0.935 & \cellcolor{green!20!yellow!35!white}0.880 & \cellcolor{green!20!yellow!35!white}0.881 & \cellcolor{green!38!yellow!35!white}0.907 & \cellcolor{green!38!yellow!35!white}0.907 \\ 
& tor\_h & \cellcolor{yellow!96!red!35!white}0.830 & \cellcolor{yellow!96!red!35!white}0.859 & \cellcolor{green!30!yellow!35!white}0.885 & \cellcolor{green!30!yellow!35!white}0.905 & & & & \\ 
\hline
\multirow{ 4}{*}{\parbox{3cm}{\texttt{Origin} (b)}} & firefox\_h & \cellcolor{green!72!yellow!35!white}0.940 & \cellcolor{green!72!yellow!35!white}0.978 & \cellcolor{green!87!yellow!35!white}0.973 & \cellcolor{green!87!yellow!35!white}0.990 & \cellcolor{green!47!yellow!35!white}0.870 & \cellcolor{green!47!yellow!35!white}0.974 & \cellcolor{green!80!yellow!35!white}0.952 & \cellcolor{green!80!yellow!35!white}0.989 \\ 
& chrome\_h & \cellcolor{green!77!yellow!35!white}0.948 & \cellcolor{green!77!yellow!35!white}0.983 & \cellcolor{green!93!yellow!35!white}0.985 & \cellcolor{green!93!yellow!35!white}0.995 & \cellcolor{green!65!yellow!35!white}0.919 & \cellcolor{green!65!yellow!35!white}0.978 & \cellcolor{green!86!yellow!35!white}0.969 & \cellcolor{green!86!yellow!35!white}0.991 \\ 
& malware\_h & \cellcolor{green!71!yellow!35!white}0.928 & \cellcolor{green!71!yellow!35!white}0.985 & \cellcolor{green!83!yellow!35!white}0.960 & \cellcolor{green!83!yellow!35!white}0.991 & \cellcolor{green!27!yellow!35!white}0.806 & \cellcolor{green!27!yellow!35!white}0.977 & \cellcolor{green!82!yellow!35!white}0.953 & \cellcolor{green!82!yellow!35!white}0.994 \\ 
& tor\_h & \cellcolor{yellow!25!red!35!white}0.520 & \cellcolor{yellow!25!red!35!white}0.957 & \cellcolor{yellow!21!red!35!white}0.510 & \cellcolor{yellow!21!red!35!white}0.955 & & & & \\ 
\hline
\hline
\multirow{ 4}{*}{\parbox{3cm}{\texttt{status-code}}} & firefox\_h & \cellcolor{green!30!yellow!35!white}0.806 & \cellcolor{green!30!yellow!35!white}0.984 & \cellcolor{green!48!yellow!35!white}0.856 & \cellcolor{green!48!yellow!35!white}0.989 & \cellcolor{green!12!yellow!35!white}0.750 & \cellcolor{green!12!yellow!35!white}0.986 & \cellcolor{green!37!yellow!35!white}0.820 & \cellcolor{green!37!yellow!35!white}0.993 \\ 
& chrome\_h & \cellcolor{green!55!yellow!35!white}0.887 & \cellcolor{green!55!yellow!35!white}0.978 & \cellcolor{green!71!yellow!35!white}0.922 & \cellcolor{green!71!yellow!35!white}0.992 & \cellcolor{green!20!yellow!35!white}0.780 & \cellcolor{green!20!yellow!35!white}0.981 & \cellcolor{green!46!yellow!35!white}0.848 & \cellcolor{green!46!yellow!35!white}0.990 \\ 
& malware\_h & \cellcolor{yellow!30!red!35!white}0.569 & \cellcolor{yellow!30!red!35!white}0.922 & \cellcolor{yellow!82!red!35!white}0.684 & \cellcolor{yellow!82!red!35!white}0.962 & \cellcolor{yellow!96!red!35!white}0.754 & \cellcolor{yellow!96!red!35!white}0.936 & \cellcolor{green!29!yellow!35!white}0.829 & \cellcolor{green!29!yellow!35!white}0.960 \\ 
& tor\_h & & & & & & & & \\ 
\hline
\multirow{ 4}{*}{\parbox{3cm}{\texttt{Content-Type}}} & firefox\_h & \cellcolor{green!12!yellow!35!white}0.817 & \cellcolor{green!12!yellow!35!white}0.919 & \cellcolor{green!23!yellow!35!white}0.848 & \cellcolor{green!23!yellow!35!white}0.923 & \cellcolor{yellow!10!red!35!white}0.652 & \cellcolor{yellow!10!red!35!white}0.778 & \cellcolor{yellow!63!red!35!white}0.766 & \cellcolor{yellow!63!red!35!white}0.825 \\ 
& chrome\_h & \cellcolor{green!38!yellow!35!white}0.875 & \cellcolor{green!38!yellow!35!white}0.940 & \cellcolor{green!58!yellow!35!white}0.919 & \cellcolor{green!58!yellow!35!white}0.957 & \cellcolor{yellow!86!red!35!white}0.777 & \cellcolor{yellow!86!red!35!white}0.882 & \cellcolor{green!32!yellow!35!white}0.880 & \cellcolor{green!32!yellow!35!white}0.917 \\ 
& malware\_h & \cellcolor{yellow!46!red!35!white}0.735 & \cellcolor{yellow!46!red!35!white}0.805 & \cellcolor{yellow!78!red!35!white}0.770 & \cellcolor{yellow!78!red!35!white}0.866 & \cellcolor{yellow!4!red!35!white}0.624 & \cellcolor{yellow!4!red!35!white}0.788 & \cellcolor{yellow!65!red!35!white}0.711 & \cellcolor{yellow!65!red!35!white}0.887 \\ 
& tor\_h & \cellcolor{yellow!0!red!35!white}0.211 & \cellcolor{yellow!0!red!35!white}0.491 & \cellcolor{yellow!0!red!35!white}0.236 & \cellcolor{yellow!0!red!35!white}0.556 & & & & \\ 
\hline
\multirow{ 4}{*}{\parbox{3cm}{\texttt{Server}}} & firefox\_h & \cellcolor{green!39!yellow!35!white}0.894 & \cellcolor{green!39!yellow!35!white}0.924 & \cellcolor{green!61!yellow!35!white}0.916 & \cellcolor{green!61!yellow!35!white}0.969 & \cellcolor{green!38!yellow!35!white}0.878 & \cellcolor{green!38!yellow!35!white}0.939 & \cellcolor{green!77!yellow!35!white}0.948 & \cellcolor{green!77!yellow!35!white}0.985 \\ 
& chrome\_h & \cellcolor{green!73!yellow!35!white}0.958 & \cellcolor{green!73!yellow!35!white}0.962 & \cellcolor{green!87!yellow!35!white}0.977 & \cellcolor{green!87!yellow!35!white}0.986 & \cellcolor{green!66!yellow!35!white}0.935 & \cellcolor{green!66!yellow!35!white}0.965 & \cellcolor{green!80!yellow!35!white}0.953 & \cellcolor{green!80!yellow!35!white}0.988 \\ 
& malware\_h & \cellcolor{yellow!87!red!35!white}0.771 & \cellcolor{yellow!87!red!35!white}0.891 & \cellcolor{green!18!yellow!35!white}0.814 & \cellcolor{green!18!yellow!35!white}0.943 & \cellcolor{green!0!yellow!35!white}0.806 & \cellcolor{green!0!yellow!35!white}0.895 & \cellcolor{green!44!yellow!35!white}0.910 & \cellcolor{green!44!yellow!35!white}0.924 \\ 
& tor\_h & \cellcolor{yellow!0!red!35!white}0.164 & \cellcolor{yellow!0!red!35!white}0.476 & \cellcolor{yellow!0!red!35!white}0.153 & \cellcolor{yellow!0!red!35!white}0.406 & & & & \\ 
\hline
\multirow{ 4}{*}{\parbox{3cm}{\texttt{Etag} (b)}} & firefox\_h & \cellcolor{green!57!yellow!35!white}0.936 & \cellcolor{green!57!yellow!35!white}0.937 & \cellcolor{green!71!yellow!35!white}0.958 & \cellcolor{green!71!yellow!35!white}0.958 & \cellcolor{yellow!92!red!35!white}0.838 & \cellcolor{yellow!92!red!35!white}0.839 & \cellcolor{green!39!yellow!35!white}0.909 & \cellcolor{green!39!yellow!35!white}0.909 \\ 
& chrome\_h & \cellcolor{green!73!yellow!35!white}0.955 & \cellcolor{green!73!yellow!35!white}0.964 & \cellcolor{green!81!yellow!35!white}0.969 & \cellcolor{green!81!yellow!35!white}0.975 & \cellcolor{green!39!yellow!35!white}0.905 & \cellcolor{green!39!yellow!35!white}0.914 & \cellcolor{green!70!yellow!35!white}0.954 & \cellcolor{green!70!yellow!35!white}0.959 \\ 
& malware\_h & \cellcolor{green!24!yellow!35!white}0.866 & \cellcolor{green!24!yellow!35!white}0.908 & \cellcolor{green!41!yellow!35!white}0.897 & \cellcolor{green!41!yellow!35!white}0.927 & \cellcolor{green!0!yellow!35!white}0.787 & \cellcolor{green!0!yellow!35!white}0.913 & \cellcolor{green!40!yellow!35!white}0.878 & \cellcolor{green!40!yellow!35!white}0.943 \\ 
& tor\_h & \cellcolor{yellow!0!red!35!white}0.606 & \cellcolor{yellow!0!red!35!white}0.651 & \cellcolor{yellow!0!red!35!white}0.676 & \cellcolor{yellow!0!red!35!white}0.703 & & & & \\ 
\hline
\multirow{ 4}{*}{\parbox{3cm}{\texttt{Via} (b)}} & firefox\_h & \cellcolor{green!75!yellow!35!white}0.962 & \cellcolor{green!75!yellow!35!white}0.965 & \cellcolor{green!83!yellow!35!white}0.975 & \cellcolor{green!83!yellow!35!white}0.976 & \cellcolor{green!42!yellow!35!white}0.892 & \cellcolor{green!42!yellow!35!white}0.936 & \cellcolor{green!65!yellow!35!white}0.934 & \cellcolor{green!65!yellow!35!white}0.961 \\ 
& chrome\_h & \cellcolor{green!81!yellow!35!white}0.958 & \cellcolor{green!81!yellow!35!white}0.985 & \cellcolor{green!83!yellow!35!white}0.964 & \cellcolor{green!83!yellow!35!white}0.987 & \cellcolor{green!59!yellow!35!white}0.918 & \cellcolor{green!59!yellow!35!white}0.959 & \cellcolor{green!79!yellow!35!white}0.960 & \cellcolor{green!79!yellow!35!white}0.979 \\ 
& malware\_h & \cellcolor{green!22!yellow!35!white}0.798 & \cellcolor{green!22!yellow!35!white}0.970 & \cellcolor{green!36!yellow!35!white}0.836 & \cellcolor{green!36!yellow!35!white}0.975 & \cellcolor{green!65!yellow!35!white}0.921 & \cellcolor{green!65!yellow!35!white}0.974 & \cellcolor{green!3!yellow!35!white}0.732 & \cellcolor{green!3!yellow!35!white}0.979 \\ 
& tor\_h & \cellcolor{yellow!0!red!35!white}0.491 & \cellcolor{yellow!0!red!35!white}0.860 & \cellcolor{yellow!2!red!35!white}0.547 & \cellcolor{yellow!2!red!35!white}0.859 & & & & \\ 
\hline
\multirow{ 4}{*}{\parbox{3cm}{\texttt{Accept-Ranges} (b)}} & firefox\_h & \cellcolor{green!63!yellow!35!white}0.946 & \cellcolor{green!63!yellow!35!white}0.946 & \cellcolor{green!70!yellow!35!white}0.956 & \cellcolor{green!70!yellow!35!white}0.956 & \cellcolor{yellow!85!red!35!white}0.825 & \cellcolor{yellow!85!red!35!white}0.831 & \cellcolor{green!40!yellow!35!white}0.909 & \cellcolor{green!40!yellow!35!white}0.911 \\ 
& chrome\_h & \cellcolor{green!75!yellow!35!white}0.959 & \cellcolor{green!75!yellow!35!white}0.969 & \cellcolor{green!85!yellow!35!white}0.975 & \cellcolor{green!85!yellow!35!white}0.980 & \cellcolor{green!35!yellow!35!white}0.901 & \cellcolor{green!35!yellow!35!white}0.904 & \cellcolor{green!69!yellow!35!white}0.954 & \cellcolor{green!69!yellow!35!white}0.956 \\ 
& malware\_h & \cellcolor{green!35!yellow!35!white}0.895 & \cellcolor{green!35!yellow!35!white}0.912 & \cellcolor{green!56!yellow!35!white}0.929 & \cellcolor{green!56!yellow!35!white}0.940 & \cellcolor{green!47!yellow!35!white}0.910 & \cellcolor{green!47!yellow!35!white}0.932 & \cellcolor{green!68!yellow!35!white}0.947 & \cellcolor{green!68!yellow!35!white}0.959 \\ 
& tor\_h & \cellcolor{yellow!0!red!35!white}0.621 & \cellcolor{yellow!0!red!35!white}0.629 & \cellcolor{yellow!0!red!35!white}0.673 & \cellcolor{yellow!0!red!35!white}0.680 & & & & \\ 
\hline
\multirow{ 4}{*}{\parbox{3cm}{\texttt{Set-Cookie} (b)}} & firefox\_h & \cellcolor{green!77!yellow!35!white}0.949 & \cellcolor{green!77!yellow!35!white}0.982 & \cellcolor{green!83!yellow!35!white}0.964 & \cellcolor{green!83!yellow!35!white}0.987 & \cellcolor{green!22!yellow!35!white}0.828 & \cellcolor{green!22!yellow!35!white}0.939 & \cellcolor{green!62!yellow!35!white}0.920 & \cellcolor{green!62!yellow!35!white}0.968 \\ 
& chrome\_h & \cellcolor{green!85!yellow!35!white}0.978 & \cellcolor{green!85!yellow!35!white}0.979 & \cellcolor{green!91!yellow!35!white}0.987 & \cellcolor{green!91!yellow!35!white}0.988 & \cellcolor{green!62!yellow!35!white}0.923 & \cellcolor{green!62!yellow!35!white}0.963 & \cellcolor{green!77!yellow!35!white}0.956 & \cellcolor{green!77!yellow!35!white}0.978 \\ 
& malware\_h & \cellcolor{green!25!yellow!35!white}0.837 & \cellcolor{green!25!yellow!35!white}0.939 & \cellcolor{green!44!yellow!35!white}0.880 & \cellcolor{green!44!yellow!35!white}0.953 & \cellcolor{green!34!yellow!35!white}0.857 & \cellcolor{green!34!yellow!35!white}0.946 & \cellcolor{green!51!yellow!35!white}0.895 & \cellcolor{green!51!yellow!35!white}0.959 \\ 
& tor\_h & \cellcolor{yellow!1!red!35!white}0.548 & \cellcolor{yellow!1!red!35!white}0.856 & \cellcolor{yellow!21!red!35!white}0.604 & \cellcolor{yellow!21!red!35!white}0.861 & & & & \\

\hline
\end{tabular}
\caption{Summary of the HTTP protocol semantics inference results.}
\label{table:summary_header_results}
\end{table*}
\egroup

\subsection{HTTP/1.1 Results}

\noindent There were a total of 72,828, 515,022, 182,498, and 50,799 HTTP/1.1 transactions in \texttt{firefox\_h}, \texttt{chrome\_h}, \texttt{malware\_h}, and \texttt{tor\_h}, respectively. This gave an average of $\sim$2.1 to $\sim$8.4 HTTP/1.1 transactions per TLS connection depending on the dataset, with Tor being a significant outlier. In these experiments, we used the first 7 days of a dataset for training and the second 7 days of the same dataset for testing. We discuss this limitation in Section \ref{sec:limitations-and-future-work} and provide additional results.

Table \ref{table:summary_header_results} provides the full set of results for each classification problem for both the initial pass and after Algorithm \ref{alg:iterative} converges. We identified TLS records containing HTTP header fields with an $F_1$ score of over 0.99 for all datasets except for \texttt{tor\_h}, which had a score of $\sim$0.87. This experiment highlights the relative difficulty that the multiplexing facilitated by the Tor protocol poses for traffic analysis relative to standalone TLS. 

Most of the other HTTP/1.1 experiments followed a similar pattern with the \texttt{tor\_h} results being significantly worse. We were able to effectively model several of the binary classification problems for the \texttt{tor\_h} dataset, with the \texttt{Cookie} and \texttt{Referer} request fields having an $F_1$ score over 0.75. The response fields performed noticeably worse due to the multiplexing behavior of Tor.

For the other datasets, we were able to achieve surprisingly competitive results across the majority of the problems. We were even able to effectively model many of the problems based on \texttt{malware\_h}, which had a much greater diversity of TLS clients and less overlap in sites visited in the training and testing datasets. Figure \ref{fig:http-confusion-matrices-primary} shows the full confusion matrices for the HTTP/1.1 response \texttt{Content-Type} header field value for \texttt{chrome\_h}, \texttt{malware\_h}, and \texttt{tor\_h}. For this problem, we achieved unweighted $F_1$ scores of 0.919, 0.770, and 0.236 for the \texttt{chrome\_h}, \texttt{malware\_h}, and \texttt{tor\_h} datasets. There was some overfitting to the majority class, \texttt{image}, which had roughly twice as many samples in each dataset than the next most represented class. Despite minor overfitting in \texttt{chrome\_h} and \texttt{malware\_h}, Figure \ref{fig:http-confusion-matrices-primary} demonstrates the feasibility of this approach to infer the value of the HTTP/1.1 response \texttt{Content-Type} header field value in an encrypted TLS tunnel. For details on the other classification problems, we refer the reader to Table \ref{table:summary_header_results}.



\begin{figure*}[t!]
\centering
\subfloat[firefox\_h] {
	\hspace{.02cm} 
   	\includegraphics[scale=0.062,valign=t]{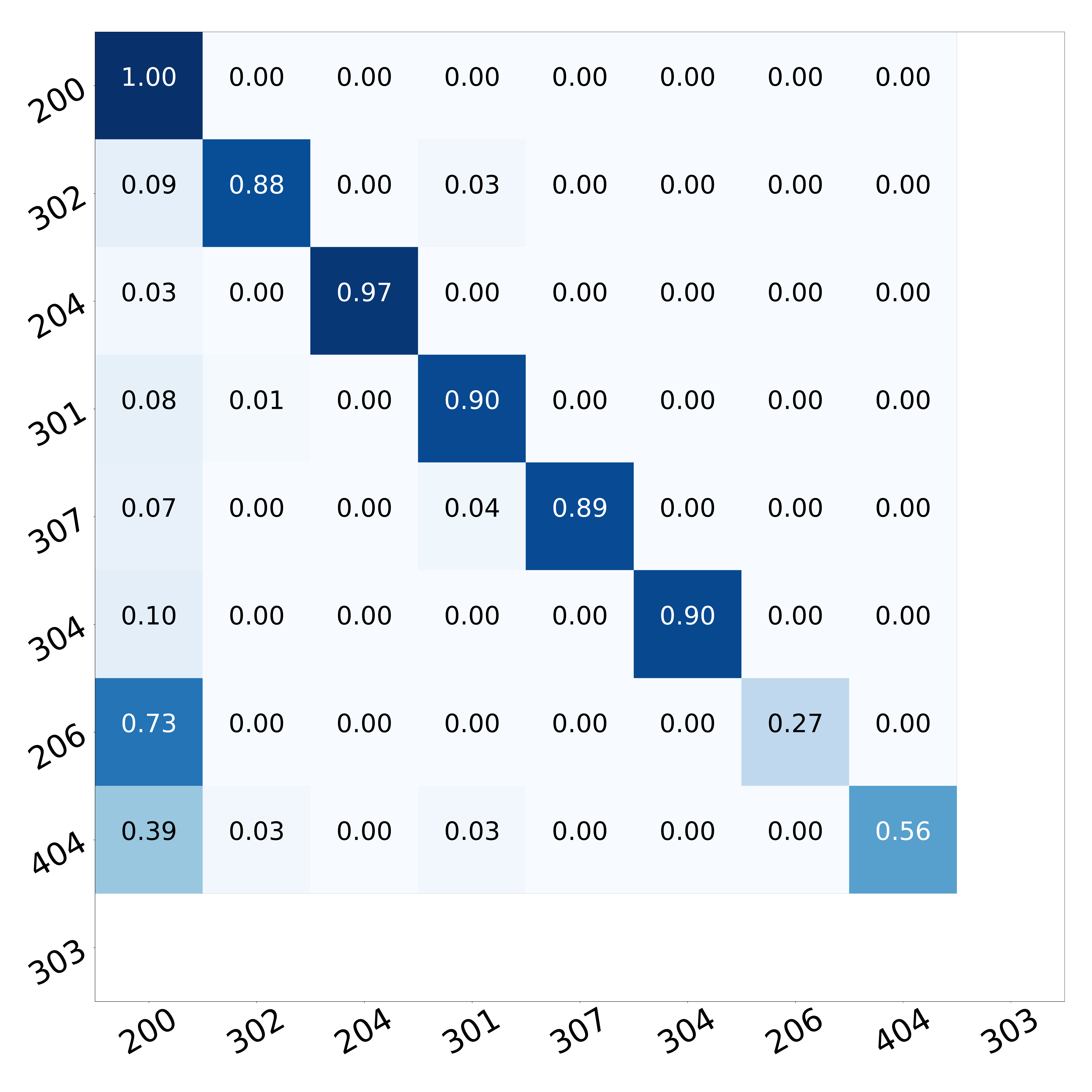}
	\label{fig:confusion-h2-firefox-code-primary}
}
\hspace{-3.75mm}
\subfloat[chrome\_h] {
   	\includegraphics[scale=0.062,valign=t]{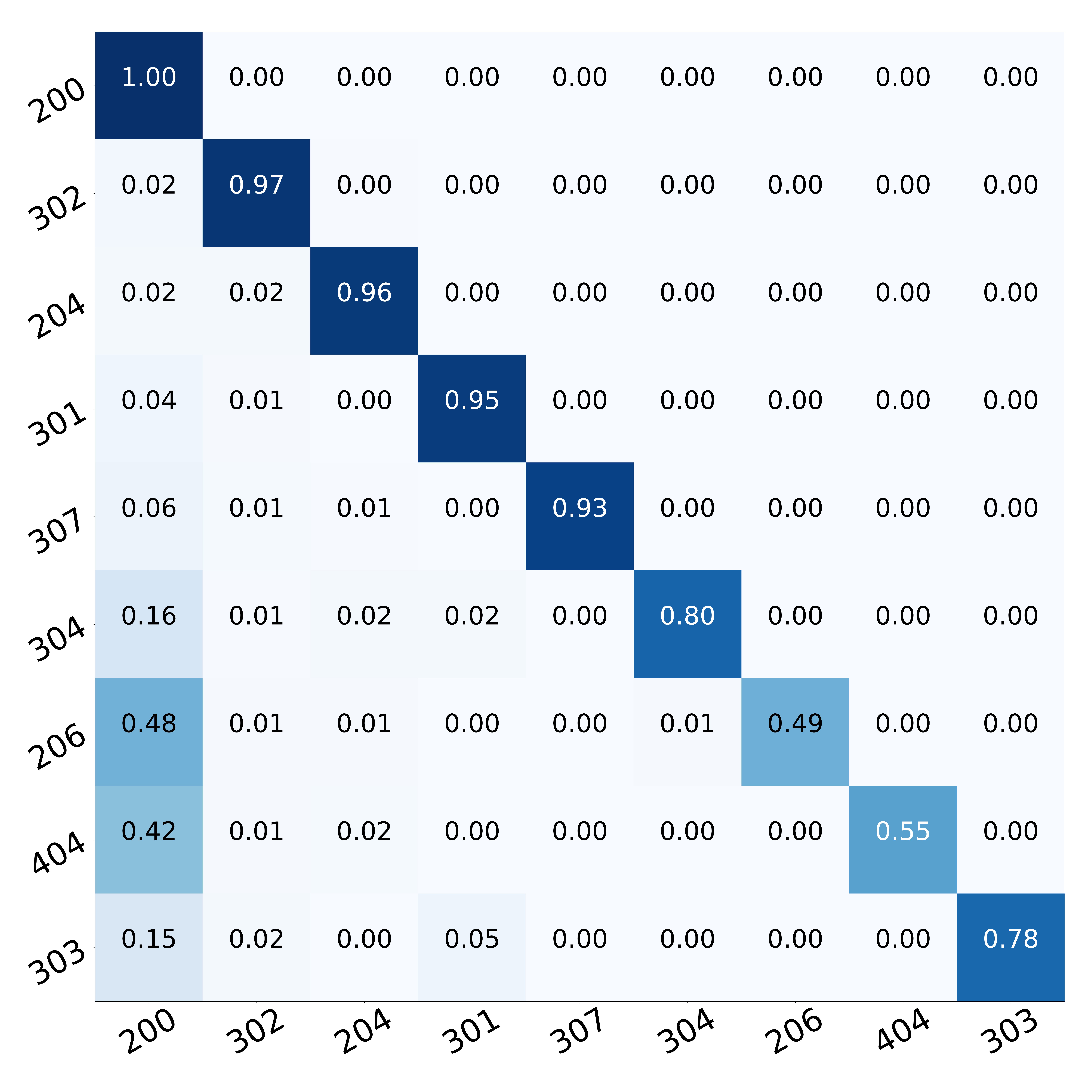}
	\label{fig:confusion-h2-chrome-code-primary}
}
\hspace{-3.75mm}
\subfloat[malware\_h] {
   	\includegraphics[scale=0.062,valign=t]{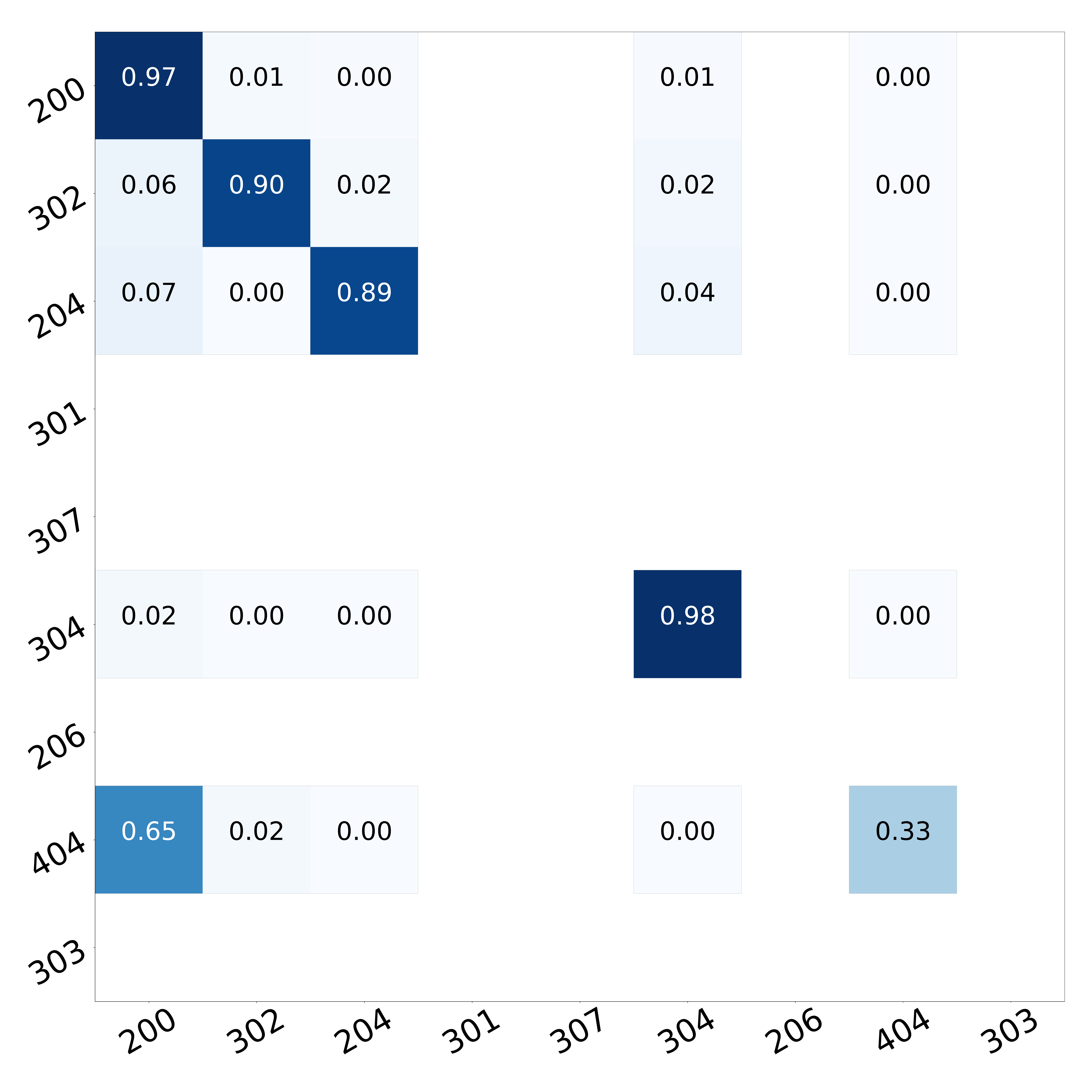}
	\label{fig:confusion-h2-malware-code-primary}
}
\caption{Confusion matrices for the HTTP/2 \texttt{status-code} value on the firefox\_h, chrome\_h, and malware\_h datasets. Matrix elements are left blank if there were no relevant labels in a particular dataset.}
\label{fig:h2-confusion-matrices-primary}
\end{figure*}

\subsection{HTTP/2 Results}

\noindent There were a total of 132,685, 561,666, and 14,734 HTTP/2 transactions in \texttt{firefox\_h}, \texttt{chrome\_h}, and \texttt{malware\_h}, respectively. This gave an average of $\sim$4 HTTP/2 transactions per TLS connection across the datasets. We performed the experiments in this section following the same structure as the HTTP/1.1 experiments. There were no HTTP/2 transactions in \texttt{tor\_h}, which was a result of the Tor Firefox process only being configured to advertise \texttt{http/1.1}.

Similar to HTTP/1.1, Table \ref{table:summary_header_results} also provides the full set of HTTP/2 results. We were able to identify TLS records containing HTTP header fields with an $F_1$ score of over 0.98 for all datasets. This slight drop in performance was expected due to the more advanced flow control mechanisms implemented by HTTP/2. In our datasets, $\sim$55\% of the TLS-HTTP/2 connections employed some form of pipelining or multiplexing. Only $\sim$15\% of the TLS-HTTP/1.1 connections employed pipelining.

The \texttt{malware\_h} HTTP/2 results were worse than the \texttt{malware\_h} HTTP/1.1 results for most problems, but we attribute this to having significantly less data in the case of HTTP/2. Both \texttt{chrome\_h} and \texttt{firefox\_h} had mostly comparable performance to the HTTP/1.1 experiments. Compared to the HTTP/1.1 results, the iterative algorithm performed exceptionally well on some problems: request \texttt{method}, request \texttt{Cookie}, request \texttt{Origin}, response \texttt{Content-Type}, and response \texttt{Server}. In these cases, the iterative algorithm was able to improve performance by effectively leveraging HTTP/2's greater number of HTTP transactions per TLS connection.

Figure \ref{fig:h2-confusion-matrices-primary} shows the confusion matrices for the HTTP/2 \texttt{status-code} header on the \texttt{firefox\_h}, \texttt{chrome\_h}, and \texttt{malware\_h} datasets, which had $F_1$ scores of 0.856, 0.922, and 0.684, respectively. Similar to other problems, the majority of the misclassifications were due to underrepresented classes being assigned to well represented classes, e.g., \texttt{206} $\rightarrow$ \texttt{200}. A more diverse and representative dataset should help mitigate these issues. The full set of results for the HTTP/2 classification problems are given in Table \ref{table:summary_header_results}.

%
%

\section{Use Cases}
\label{sec:use_cases}

\noindent We now examine two possible applications of our techniques: improved malware detection and website fingerprinting. Our goal in this section is to test the feasibility of using the inferences introduced in this paper to improve the performance of these use cases; we did not attempt to demonstrate hyper-optimized results. We used the full two weeks of the previous datasets to train the classifiers needed to perform our iterative HTTP protocol semantics inferences. We then used the trained classifiers and Algorithm \ref{alg:iterative} to enrich the samples related to the two use cases. If available, we did \textbf{not} make use of any decrypted data features for samples in this section. \texttt{firefox\_h}, \texttt{chrome\_h}, and \texttt{malware\_h} were used to train the classifiers needed for Section \ref{sec:malware-detection}, and \texttt{tor\_h} was used to train the classifiers needed for Section \ref{sec:website-fingerprinting}.

\subsection{Malware Detection}
\label{sec:malware-detection}

\bgroup
\def\arraystretch{1.2}
\begin{table}[t!]\small
\center
\begin{tabular}{|l|c|c|c|c|}
\hline
Feature Set & $F_1$ Score & Precision & Recall & Acc \\
\hline
Standard & 0.951 & 0.951 & 0.915 & 0.958 \\
Enriched & 0.979 & 0.984 & 0.959 & 0.982 \\
\hline
\end{tabular}
\caption{Malware classification results using a standard feature set and an enriched feature set that takes advantage of HTTP protocol semantics inferences.}
\label{table:malware-classification-results}
\end{table}
\egroup

\noindent As described in Section \ref{section:malware-data}, we used \texttt{enterprise\_m} and \texttt{malware\_m} from Table \ref{table:dataset_summary} to test if first inferring the semantics of encrypted HTTP transactions can improve malware detection. We used the November data for training and the December data for testing. We explored two feature sets for this problem. The standard feature set included the 108 connection-dependent features described in Section \ref{sec:http-header-data-features}. In the standard set, we also used TLS-specific features: 
\begin{enumerate}[itemsep=-1mm]
\item Binary features for the 100 most commonly offered cipher suites
\item Binary features for the 25 most commonly advertised TLS extensions (including GREASE extensions  \cite{grease}, which are treated as a single feature)
\item A categorical feature for the selected cipher suite
\end{enumerate} 
There were 234 total features for the standard set. The enhanced set included all 234 features of the standard set, and the features representing the predicted values from Algorithm \ref{alg:iterative} and described in Section \ref{sec:iterative-classification}. In total, there were either 302 features for HTTP/1.1 TLS connections or 308 features for HTTP/2 TLS connections. We trained a standard random forest model as described in Section \ref{sec:header-inferences} to classify the test samples.

As Table \ref{table:malware-classification-results} demonstrates, applying the iterative HTTP protocol semantics classifier and learning an intermediate representation of the HTTP transactions within the encrypted TLS tunnel significantly helped the performance of the classifier. The header inferences increased the $F_1$ score from 0.951 to 0.979, and had similar impacts to precision and recall. These results are a notable improvement over previous results on malware TLS detection \cite{anderson16deciphering}, which relied on additional data features such as server certificates to obtain higher accuracy. Because TLS 1.3 obfuscates certificates \cite{tls13}, new techniques will be needed to address this use case.

Table \ref{table:malware-feature-importances} lists the 10 most importance features for classification in the standard and enriched feature sets. This ranking was generated by computing the Gini feature importance \cite{breiman1984classification}. From standard feature set, the first TLS record length corresponding to the \texttt{client\_hello} was informative. The 7th and 8th TLS record lengths were also informative because the enterprise dataset contained more examples of session resumption. From the enriched feature set, HTTP fields aimed at tracking and keeping state were among the most useful, e.g., \texttt{Set-Cookie} and \texttt{Referer}. The \texttt{method} and \texttt{status-code} were also in the top 10 due to malware being more likely to perform a \texttt{GET} request with a resulting \texttt{404} status code.

The performance of the enriched feature set would still yield too many false positives on a real network, but it is only looking at data features related to a single TLS session. Our techniques could easily be incorporated into a more comprehensive network monitoring architecture that correlates multiple connections and independent data sources.

\bgroup
\def\arraystretch{1.2}
\begin{table}[t!]\small
\center
\begin{tabular}{|c|l|l|}
\hline
Rank & Standard & Enriched \\
\hline
\hline
 1 & 8th Record Length & 8th Record Length \\
\hline
 2 & 1st Record Length & HTTP: \texttt{Set-Cookie} \\
\hline
 3 & \# Out Bytes             & \# Out Bytes \\
\hline
 4 & \# Out Records           & 1st Record Length \\
\hline
 5 & Offered Cipher Suite: & HTTP: \texttt{Referer} \\
   & \texttt{DHE\_DSS\_WITH\_}   & \\
   & \texttt{AES\_256\_CBC\_SHA}   & \\
\hline
 6 & \# In Records & HTTP: \texttt{Content-Type} \\
\hline
 7 & \# In Bytes & \# In Records \\
\hline
 8 & Advertised Extension: & HTTP: \texttt{status-code} \\
   & \texttt{channel\_id}   & \\
\hline
 9 & Duration & \# Out Packets \\
\hline
 10 & 7th Record Length & HTTP: \texttt{method} \\
\hline
\end{tabular}
\caption{The 10 most important features for classifying malware with the standard and enhanced features sets.}
\label{table:malware-feature-importances}
\end{table}
\egroup

\subsection{Website Fingerprinting}
\label{sec:website-fingerprinting}

\noindent We used \texttt{tor\_open\_w} and \texttt{tor\_censor\_w} from Table \ref{table:dataset_summary} in our website fingerprinting experiments. Similar to previous work on website fingerprinting \cite{wang2014effective}, \texttt{tor\_censor\_w} contained 50 samples per monitored site from a list of 50 sites currently blocked in some countries pertaining to information dissemination (e.g., \texttt{twitter.com}), covert communication (e.g., \texttt{torproject.org}), and pornographic imagery. \texttt{tor\_open\_w} contained 5,000 non-monitored samples, where each sample is a unique connection to a website in the Alexa top-10k. These sites were contacted in order, and we selected the first 5,000 sites that properly decrypted and did not have any HTTP requests to a monitored site.

The feature set in this experiment was based on the features used by Wang et al. \cite{wang2014effective}, and included total packet count, unique packet lengths, several counts related to burst patterns, and the lengths of the first 20 packets. A more detailed description of these features is provided by Wang et al. \cite{wang2014effective}. We took advantage of Wang et al.'s $k$-NN classifier and hyperparameter learning algorithms with $k_{reco}$ fixed to the suggested value of 5 \cite{wang2014effective}. We used 10-fold CV, ensuring non-monitored sites were not in the training and testing sets simultaneously. 

\begin{figure}
	\centering 
   \includegraphics[scale=0.51]{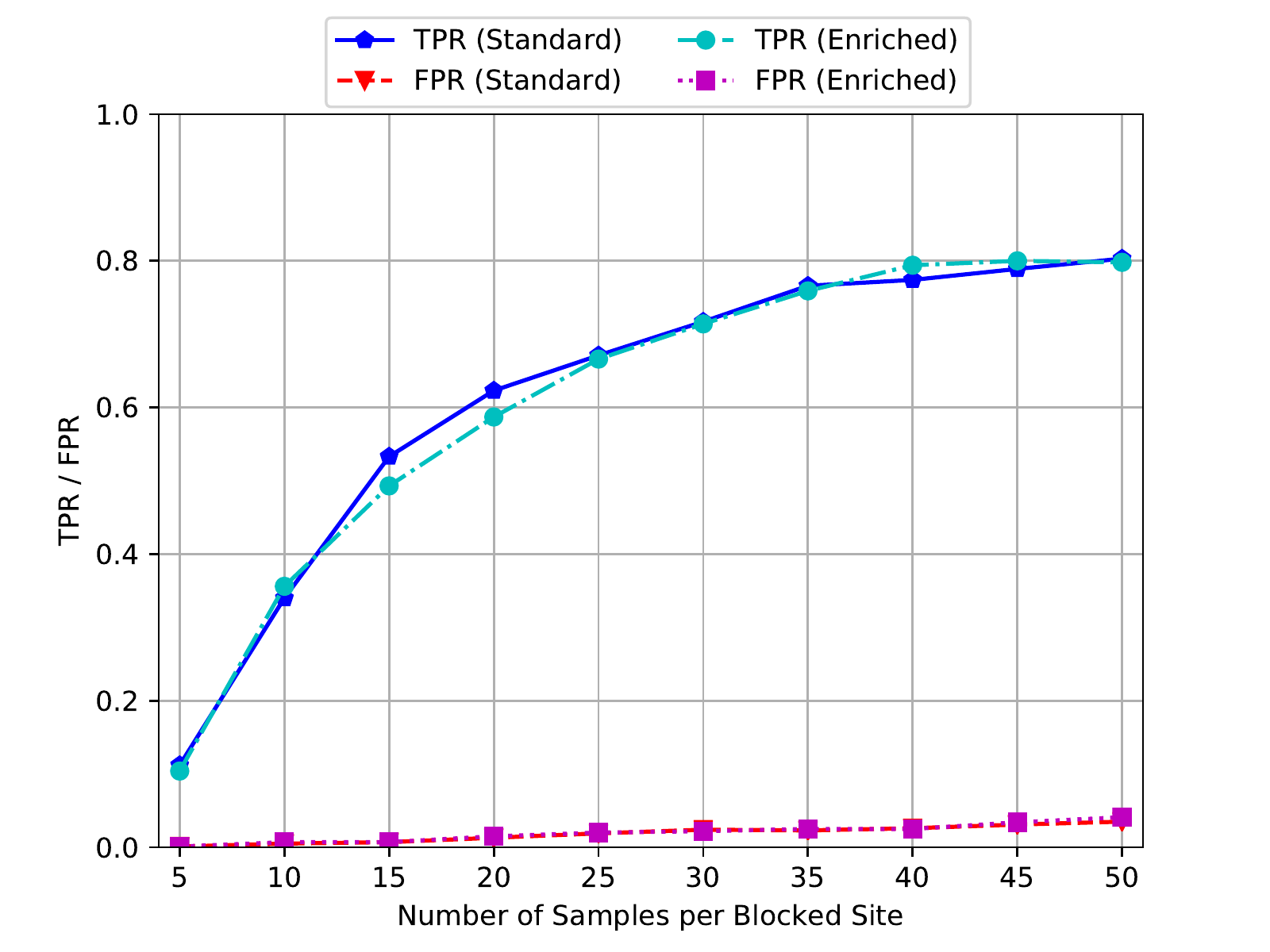}
	\caption{Website fingerprinting learning curve as the number of samples per blocked site is adjusted.}
	\label{fig:website-fingerprinting-learning-curve}
\end{figure}

Figure \ref{fig:website-fingerprinting-learning-curve} provides the results of this experiment as we adjust the number of unique samples per monitored site from 5 to 50 in increments of 5. The introduction of the header inferences seem to be strictly adding noise that is effectively filtered out during the weight adjustment algorithm of Wang et al. In Section \ref{sec:limitations-and-future-work}, we provide some references that may increase the power of this attack.

\section{Discussion}
\label{sec:limitations-and-future-work}

\noindent We have shown that it is possible to infer HTTP protocol semantics within encrypted TLS tunnels, but our results depend on the existence of a rich source of labeled data correlating the encrypted traffic patterns observable on a network with the underlying HTTP transactions. The results based on \texttt{malware\_h} were the closest to a real-world scenario where the testing dataset consisted of many different clients visiting servers unseen in training. Table \ref{table:chrome_firefox_results} shows the performance of our algorithm when training on \texttt{chrome\_h} and testing on \texttt{firefox\_h}, and as expected, the results are worse. One observation we made was the difference by the two browsers in utilizing HTTP/1.1 pipelining: 10\% of TLS-HTTP/1.1 Chrome connections used pipelining versus 25\% for Firefox. In many of these cases, Chrome would create multiple TLS connections instead of pipelining the requests. There are undoubtedly other differences that would cause the models to not easily transfer between browsers and even operating systems. A deployable system would need to first recognize the operating system and application, which can be accomplished by readily available tools \cite{brotherston2015tls,p0f2012tls}, and then apply the relevant HTTP protocol semantics models to the data features extracted from the encrypted tunnel. This would require curating an extensive dataset, which is not an unreasonable burden for a well-funded department.

Our results indicated that the Tor protocol provides a suitable defense for entities wishing to defend against the methods of this paper. The fixed-length Tor cells and multiplexing many distinct TLS connections over a single TLS-Tor connection reduces the signal in many of our chosen features. This can be seen to a lesser extent in Table \ref{table:summary_header_results} with respect to HTTP/2 over TLS due to HTTP/2's greater use of pipelining and multiplexing in our datasets. Multiplexing communication to different origins over a single HTTP/2 connection has been recently proposed \cite{http2cert, http2origin}, which would most likely degrade our results even further. Wang and Goldberg put forward several algorithms to properly segment and de-noise Tor streams \cite{wang2016realistically}. Preprocessing \texttt{tor\_h} with Wang and Goldberg's methods may increase the efficacy of our algorithms, and their techniques may also have applications to future implementations of HTTP/2. These investigations are left for future work.

\bgroup
\def\arraystretch{1.2}
\begin{table}[t!]\small
\center
\begin{tabular}{|l|c|c|}
\hline
Problem & $F_1$ Score & Accuracy \\
\hline
method        & 0.717 & 0.971 \\
Content-Type  & 0.790 & 0.841 \\
\hline
status-code   & 0.492 & 0.869 \\
Content-Type  & 0.400 & 0.421 \\
Server        & 0.741 & 0.830 \\
\hline
\end{tabular}
\caption{HTTP/1.1 semantics inference results when training with \texttt{chrome\_h} and testing with \texttt{firefox\_h}.}
\label{table:chrome_firefox_results}
\end{table}
\egroup

In the context of malware detection, our results on Tor demonstrated that malware employing common evasion strategies \cite{tor, luo2011httpos, wright2009traffic} should be able to mislead our HTTP protocol semantics inferences, resulting in the addition of noisy data features to the malware classifier. These evasion strategies can be detected \cite{wang15seeing}. Although these detection methods are not specific to malware communication, they could be used to inform the malware classifier to avoid using the inferred HTTP data features. From an adversarial point-of-view, it is also important to note that the feature importances given in Table \ref{table:malware-feature-importances} are not static, and the relative importance of any particular feature is likely to change over time. However, with continual training, malware classifiers will be able to take advantage of the insights provided by the HTTP inferences presented in this paper.

Our paper's intended purpose is to highlight a fundamental, novel framework that has many potential applications. While we leave developing these applications to future work, we believe our techniques can be applied to de-anonymizing a NAT'd user's browsing history by using the presence of the \texttt{Referer}/\texttt{Origin}/\texttt{Cookie} fields together with timestamps. Our techniques can also identify transmitted file sizes at a much higher resolution by annotating each TLS record with the HTTP/2 frame types it contains, and discarding TLS records that do not contain relevant \texttt{DATA} frames. An organization could leverage a known set of sensitive file sizes correlated with endpoints, and use the methods presented in this paper to identify out-of-policy file movements. Finally, for maintenance and debugging of servers on a network, our method is superior to active scanning, which can be inefficient and ineffective due to limited client configurations and imperfect information about the client's request and server's response. The HTTP inference techniques presented in this paper can identify problems without relying on active scanning or TLS termination proxies.

\section{Ethical Considerations}

\noindent The majority of our data was collected in a lab environment and did not contain sensitive information. The Tor data collection adhered to the ethical Tor research guidelines \cite{ethical-tor}. The data collection for the malware detection experiments did contain highly confidential and sensitive data. We followed all institutional procedures, and obtained the appropriate authorizations.  While collecting the data, all IP addresses and enterprise user names were anonymized via deterministic encryption.

\section{Conclusions}

\noindent In this paper, we have shown that it is possible to infer many of the underlying HTTP protocol features without needing to compromise the encryption that secures the HTTPS protocol. Our framework can correctly identify HTTP/1.1 and HTTP/2 records transmitted over HTTPS with $F_1$ scores greater than 0.99 and 0.98, respectively. Once the HTTP records are identified, our system uses multi-class classification to identify the value of several fields, e.g., \texttt{Content-Type}, and binary classification to identify the presence or absence of additional fields, e.g., \texttt{Cookie}. We have demonstrated competitive results on datasets composed of hundreds-of-thousands of encrypted HTTP transactions taken from Firefox 58.0, Chrome 63.0, and a malware analysis sandbox. We also applied our techniques to Tor Browser 7.0.11, but achieved significantly lower accuracies, suggesting that the Tor protocol is robust against these methods.

Inferences on the semantics of HTTP have intrinsic value that can be used by both attackers and defenders. For example, network administrators can use these inferences to passively monitor dynamic, complex networks to ensure a proper security posture and perform debugging without the need of TLS termination proxies, and attackers can use these inferences to  de-anonymize a NAT'd user's browsing history. We performed two experiments that highlight this moral tension: leveraging encrypted HTTP inferences to improve malware detection and website fingerprinting over Tor. We showed that our techniques can improve the detection of encrypted malware communication, but they failed to improve website fingerprinting due to the defense mechanisms implemented by the Tor protocol. Given our broader set of results and the increasing sophistication of network traffic analysis techniques, future research is needed to evaluate the confidentiality goals of TLS with respect to users' expectations of privacy.



{\footnotesize
\bibliography{limitless}

\begin{thebibliography}{40}
\providecommand{\natexlab}[1]{#1}
\providecommand{\url}[1]{\texttt{#1}}
\expandafter\ifx\csname urlstyle\endcsname\relax
  \providecommand{\doi}[1]{doi: #1}\else
  \providecommand{\doi}{doi: \begingroup \urlstyle{rm}\Url}\fi

\bibitem[Anderson and McGrew(2016)]{anderson16identifying}
B.~Anderson and D.~McGrew.
\newblock Identifying {E}ncrypted {M}alware {T}raffic with {C}ontextual {F}low
  {D}ata.
\newblock In \emph{ACM Workshop on Artificial Intelligence and Security
  (AISec)}, pages 35--46, 2016.

\bibitem[Anderson and McGrew(2017)]{anderson2017noisy}
B.~Anderson and D.~McGrew.
\newblock {Machine Learning for Encrypted Malware Traffic Classification:
  Accounting for Noisy Labels and Non-Stationarity}.
\newblock In \emph{ACM SIGKDD International Conference on Knowledge Discovery
  in Data Mining (KDD)}, pages 1723--1732, 2017.

\bibitem[{Anderson} et~al.(2017){Anderson}, {Paul}, and
  {McGrew}]{anderson16deciphering}
B.~{Anderson}, S.~{Paul}, and D.~{McGrew}.
\newblock {Deciphering {M}alware's {U}se of {TLS} (without {D}ecryption)}.
\newblock \emph{Journal of Computer Virology and Hacking Techniques}, pages
  1--17, 2017.

\bibitem[Belshe et~al.(2015)Belshe, Peon, and Thomson]{http2}
M.~Belshe, R.~Peon, and M.~Thomson.
\newblock Hypertext {T}ransfer {P}rotocol {V}ersion 2 ({HTTP}/2).
\newblock RFC 7540 (Proposed Standard), 2015.
\newblock \url{http://www.ietf.org/rfc/rfc7540.txt}.

\bibitem[Benjamin(2017)]{grease}
D.~Benjamin.
\newblock {Applying GREASE to TLS Extensibility}.
\newblock Internet-Draft (Informational), 2017.
\newblock \url{https://www.ietf.org/archive/id/draft-ietf-tls-grease-00.txt}.

\bibitem[Bishop et~al.(2017)Bishop, Sullivan, and Thomson]{http2cert}
M.~Bishop, N.~Sullivan, and M.~Thomson.
\newblock {Secondary Certificate Authentication in HTTP/2}.
\newblock Internet-Draft (Standards Track), 2017.
\newblock
  \url{https://tools.ietf.org/html/draft-bishop-httpbis-http2-additional-certs-05}.

\bibitem[Breiman et~al.(1984)Breiman, Friedman, Stone, and
  Olshen]{breiman1984classification}
L.~Breiman, J.~Friedman, C.~J. Stone, and R.~A. Olshen.
\newblock \emph{Classification and {R}egression {T}rees}.
\newblock CRC press, 1984.

\bibitem[Brotherston(2015)]{brotherston2015tls}
L.~Brotherston.
\newblock {Stealthier Attacks and Smarter Defending with TLS Fingerprinting}.
\newblock \emph{DerbyCon}, 2015.

\bibitem[Dierks and Rescorla(2008)]{tls12}
T.~Dierks and E.~Rescorla.
\newblock The {T}ransport {L}ayer {S}ecurity ({TLS}) {P}rotocol {V}ersion 1.2.
\newblock RFC 5246 (Proposed Standard), 2008.
\newblock \url{http://www.ietf.org/rfc/rfc5246.txt}.

\bibitem[Dingledine and Mathewson(2017)]{tor}
R.~Dingledine and N.~Mathewson.
\newblock Tor {P}rotocol {S}pecification.
\newblock \url{https://gitweb.torproject.org/torspec.git/tree/tor-spec.txt},
  2017.

\bibitem[Durumeric et~al.(2017)Durumeric, Ma, Springall, Barnes, Sullivan,
  Bursztein, Bailey, Halderman, and Paxson]{httpsInterception2017}
Z.~Durumeric, Z.~Ma, D.~Springall, R.~Barnes, N.~Sullivan, E.~Bursztein,
  M.~Bailey, J.~A. Halderman, and V.~Paxson.
\newblock {The Security Impact of HTTPS Interception}.
\newblock In \emph{Network and Distributed System Security Symposium (NDSS)},
  2017.

\bibitem[Fielding and Reschke(2014{\natexlab{a}})]{http11_message_syntax}
R.~Fielding and J.~Reschke.
\newblock Hypertext {T}ransfer {P}rotocol ({HTTP}/1.1): {M}essage {S}yntax and
  {R}outing.
\newblock RFC 7230 (Proposed Standard), 2014{\natexlab{a}}.
\newblock \url{http://www.ietf.org/rfc/rfc7230.txt}.

\bibitem[Fielding and Reschke(2014{\natexlab{b}})]{http11_semantics_content}
R.~Fielding and J.~Reschke.
\newblock Hypertext {T}ransfer {P}rotocol ({HTTP}/1.1): {S}emantics and
  {C}ontent.
\newblock RFC 7231 (Proposed Standard), 2014{\natexlab{b}}.
\newblock \url{http://www.ietf.org/rfc/rfc7231.txt}.

\bibitem[Goldwasser and Micali(1982)]{Goldwasser:1982:PEA:800070.802212}
S.~Goldwasser and S.~Micali.
\newblock {Probabilistic Encryption \& How to Play Mental Poker Keeping Secret
  All Partial Information}.
\newblock In \emph{ACM Symposium on Theory of Computing (STOC)}, pages
  365--377. ACM, 1982.

\bibitem[Goldwasser and Micali(1984)]{DBLP:journals/jcss/GoldwasserM84}
S.~Goldwasser and S.~Micali.
\newblock {Probabilistic Encryption}.
\newblock \emph{J. Comput. Syst. Sci.}, 28\penalty0 (2):\penalty0 270--299,
  1984.

\bibitem[Green et~al.(2018)Green, Droms, Housley, Turner, and
  Fenter]{draft-green-tls-static-dh-in-tls13-01}
M.~Green, R.~Droms, R.~Housley, P.~Turner, and S.~Fenter.
\newblock {Data Center use of Static Diffie-Hellman in TLS 1.3}.
\newblock Work in Progress, July 2018.
\newblock URL
  \url{\url{https://tools.ietf.org/id/draft-green-tls-static-dh-in-tls13-01.txt}}.

\bibitem[Gu et~al.(2008)Gu, Perdisci, Zhang, and Lee]{gu2008botminer}
G.~Gu, R.~Perdisci, J.~Zhang, and W.~Lee.
\newblock Bot{M}iner: {C}lustering {A}nalysis of {N}etwork {T}raffic for
  {P}rotocol-and {S}tructure-{I}ndependent {B}otnet {D}etection.
\newblock In \emph{USENIX Security Symposium}, pages 139--154, 2008.

\bibitem[Halderman et~al.(2009)Halderman, Schoen, Heninger, Clarkson, Paul,
  Calandrino, Feldman, Appelbaum, and Felten]{halderman09:coldboot}
J.~A. Halderman, S.~D. Schoen, N.~Heninger, W.~Clarkson, W.~Paul, J.~A.
  Calandrino, A.~J. Feldman, J.~Appelbaum, and E.~W. Felten.
\newblock {Lest We Remember: Cold-Boot Attacks on Encryption Keys}.
\newblock \emph{Communications of the ACM}, 52\penalty0 (5):\penalty0 91--98,
  2009.

\bibitem[Kambic(2016)]{kambic16:schannel}
J.~Kambic.
\newblock Cunning with {CNG}: Soliciting secrets from {Schannel}.
\newblock Black Hat USA, 2016.

\bibitem[Liberatore and Levine(2006)]{liberatore2006inferring}
M.~Liberatore and B.~N. Levine.
\newblock {Inferring the Source of Encrypted HTTP Connections}.
\newblock In \emph{Proceedings of the Thirteenth ACM Conference on Computer and
  Communications Security (CCS)}, pages 255--263, 2006.

\bibitem[Ligh et~al.(2014)Ligh, Case, Levy, and Walters]{ligh14:artofmem}
M.~H. Ligh, A.~Case, J.~Levy, and A.~Walters.
\newblock \emph{{The Art of Memory Forensics: Detecting Malware and Threats in
  Windows, Linux, and Mac Memory}}.
\newblock John Wiley \& Sons, 2014.

\bibitem[Luo et~al.(2011)Luo, Zhou, Chan, Lee, Chang, and
  Perdisci]{luo2011httpos}
X.~Luo, P.~Zhou, E.~W. Chan, W.~Lee, R.~K. Chang, and R.~Perdisci.
\newblock {HTTPOS: Sealing Information Leaks with Browser-Side Obfuscation of
  Encrypted Flows}.
\newblock In \emph{Network and Distributed System Security Symposium (NDSS)},
  2011.

\bibitem[Majkowski()]{p0f2012tls}
M.~Majkowski.
\newblock {SSL Fingerprinting for p0f}.
\newblock
  \url{https://idea.popcount.org/2012-06-17-ssl-fingerprinting-for-p0f/}.

\bibitem[Nottingham and Nygren(2017)]{http2origin}
M.~Nottingham and E.~Nygren.
\newblock {The ORIGIN HTTP/2 Frame}.
\newblock Internet-Draft (Standards Track), 2017.
\newblock \url{https://tools.ietf.org/html/draft-ietf-httpbis-origin-frame-06}.

\bibitem[Oh et~al.(2017)Oh, Li, and Hopper]{oh2017fingerprinting}
S.~E. Oh, S.~Li, and N.~Hopper.
\newblock {Fingerprinting Keywords in Search Queries over Tor}.
\newblock \emph{Proceedings of Privacy Enhancing Technologies (PETS)},
  2017:\penalty0 251--270, 2017.

\bibitem[Panchenko et~al.(2011)Panchenko, Niessen, Zinnen, and
  Engel]{panchenko2011website}
A.~Panchenko, L.~Niessen, A.~Zinnen, and T.~Engel.
\newblock {Website Fingerprinting in Onion Routing Based Anonymization
  Networks}.
\newblock In \emph{Proceedings of the Tenth annual ACM Workshop on Privacy in
  the Electronic Society (WPES)}, pages 103--114, 2011.

\bibitem[Panchenko et~al.(2016)Panchenko, Lanze, Pennekamp, Engel, Zinnen,
  Henze, and Wehrle]{panchenko2016website}
A.~Panchenko, F.~Lanze, J.~Pennekamp, T.~Engel, A.~Zinnen, M.~Henze, and
  K.~Wehrle.
\newblock {Website Fingerprinting at Internet Scale}.
\newblock In \emph{Network and Distributed System Security Symposium (NDSS)},
  2016.

\bibitem[Peon and Ruellan(2015)]{hpack}
R.~Peon and H.~Ruellan.
\newblock {HPACK: Header Compression for HTTP/2}.
\newblock RFC 7541 (Proposed Standard), 2015.
\newblock \url{http://www.ietf.org/rfc/rfc7541.txt}.

\bibitem[Reed and Kranch(2017)]{reed2017identifying}
A.~Reed and M.~Kranch.
\newblock Identifying {HTTPS}-{P}rotected {N}etflix {V}ideos in {R}eal-{T}ime.
\newblock In \emph{Proceedings of the Seventh ACM on Conference on Data and
  Application Security and Privacy (CODASPY)}, pages 361--368, 2017.

\bibitem[Rescorla(2017)]{tls13}
E.~Rescorla.
\newblock The {T}ransport {L}ayer {S}ecurity ({TLS}) {P}rotocol {V}ersion 1.3
  (draft 23).
\newblock Intended Status: Standards Track, 2017.
\newblock \url{https://tools.ietf.org/html/draft-ietf-tls-tls13-23}.

\bibitem[Schuster et~al.(2017)Schuster, Shmatikov, and
  Tromer]{schuster2017beauty}
R.~Schuster, V.~Shmatikov, and E.~Tromer.
\newblock Beauty and the {B}urst: {R}emote {I}dentification of {E}ncrypted
  {V}ideo {S}treams.
\newblock pages 1357--1374, 2017.

\bibitem[Tegeler et~al.(2012)Tegeler, Fu, Vigna, and
  Kruegel]{tegeler2012botfinder}
F.~Tegeler, X.~Fu, G.~Vigna, and C.~Kruegel.
\newblock Botfinder: {F}inding {B}ots in {N}etwork {T}raffic without {D}eep
  {P}acket {I}nspection.
\newblock In \emph{ACM International Conference on Emerging Networking
  Experiments and Technologies (Co-NEXT)}, pages 349--360, 2012.

\bibitem[{The Tor Project}()]{ethical-tor}
{The Tor Project}.
\newblock Ethical {T}or {R}esearch: {G}uidelines.
\newblock \url{https://blog.torproject.org/ethical-tor-research-guidelines}.
\newblock Accessed: 2018-01-15.

\bibitem[US-CERT(2017)]{uscert2017https}
US-CERT.
\newblock {HTTPS Interception Weakens TLS Security}.
\newblock \url{https://www.us-cert.gov/ncas/alerts/TA17-075A}, 2017.

\bibitem[Van~Goethem et~al.(2016)Van~Goethem, Vanhoef, Piessens, and
  Joosen]{van2016request}
T.~Van~Goethem, M.~Vanhoef, F.~Piessens, and W.~Joosen.
\newblock {Request and Conquer: Exposing Cross-Origin Resource Size}.
\newblock In \emph{USENIX Security Symposium}, pages 447--462, 2016.

\bibitem[Wagner and Schneier(1996)]{wagner1996analysis}
D.~Wagner and B.~Schneier.
\newblock {Analysis of the SSL 3.0 protocol}.
\newblock In \emph{The Second USENIX Workshop on Electronic Commerce
  Proceedings}, pages 29--40, 1996.

\bibitem[Wang et~al.(2015)Wang, Dyer, Akella, Ristenpart, and
  Shrimpton]{wang15seeing}
L.~Wang, K.~P. Dyer, A.~Akella, T.~Ristenpart, and T.~Shrimpton.
\newblock {Seeing Through Network-Protocol Obfuscation}.
\newblock In \emph{Proceedings of the Twenty-Second ACM Conference on Computer
  and Communications Security (CCS)}, pages 57--69, 2015.

\bibitem[Wang and Goldberg(2016)]{wang2016realistically}
T.~Wang and I.~Goldberg.
\newblock On {R}ealistically {A}ttacking {T}or with {W}ebsite {F}ingerprinting.
\newblock \emph{Proceedings of Privacy Enhancing Technologies (PETS)}, pages
  21--36, 2016.

\bibitem[Wang et~al.(2014)Wang, Cai, Nithyanand, Johnson, and
  Goldberg]{wang2014effective}
T.~Wang, X.~Cai, R.~Nithyanand, R.~Johnson, and I.~Goldberg.
\newblock Effective {A}ttacks and {P}rovable {D}efenses for {W}ebsite
  {F}ingerprinting.
\newblock In \emph{USENIX Security Symposium}, pages 143--157, 2014.

\bibitem[Wright et~al.(2009)Wright, Coull, and Monrose]{wright2009traffic}
C.~Wright, S.~Coull, and F.~Monrose.
\newblock {Traffic Morphing: An Efficient Defense Against Statistical Traffic
  Analysis}.
\newblock In \emph{Network and Distributed System Security Symposium (NDSS)},
  2009.

\end{thebibliography}
}

\end{document}